# Roadmap: Emerging Platforms and Applications of Optical Frequency Combs and Dissipative Solitons

**Abstract:** The discovery of optical frequency combs (OFCs) has revolutionised science and technology by bridging electronics and photonics, driving major advances in precision measurements, atomic clocks, spectroscopy, telecommunications, and astronomy. However, current OFC systems still require further development to enable broader adoption in fields such as communication, aerospace, defence, and healthcare. There is a growing need for compact, portable OFCs that deliver high output power, robust self-referencing, and application-specific spectral coverage. On the conceptual side, progress toward such systems is hindered by an incomplete understanding of the fundamental principles governing OFC generation in emerging devices and materials, as well as evolving insights into the interplay between soliton and mode-locking effects. This roadmap presents the vision of a diverse group of academic and industry researchers and educators from Europe, along with their collaborators, on the current status and future directions of OFC science. It highlights a multidisciplinary approach that integrates novel physics, engineering innovation, and advanced researcher training. Topics include advances in soliton science as it relates to OFCs, the extension of OFC spectra into the visible and mid-infrared ranges, metrology applications and noise performance of integrated OFC sources, new fibre-based OFC modules, OFC lasers and OFC applications in astronomy.

**List of Contributions:**
1) Towards frequency agile solitons in microresonators,
*Dmitry Skryabin*
2) Optical Microcombs for Precision Metrology Applications,
*Arne Kordts, Richard Zeltner, and Ronald Holzwarth*
3) Ultimate coherence performance limits of soliton microcombs,
*Victor Torres-Company, Tobias Herr, Fuchuan Lei, and Qi-Fan Yang*
4) Kerr combs in microresonators with linearly coupled modes,
*Tobias Herr, Erwan Lucas, and Victor Torres Company*
5) Integrated resonant up-conversion of Kerr frequency comb,
*Camille-Sophie Brès*
6) Soliton dynamics in dual-mode and dual-pump microresonators,
*John F. Donegan and Hai-Zhong Weng*
7) Towards mid-IR frequency comb sources in passive photonic platforms,
*Delphine Marris-Morini and Adel Bousseksou*
8) Frequency combs produced by femtosecond optical parametric generators and oscillators,
*Markku Vainio*
9) Frequency combs in high-quality-factor fibre Fabry-Pérot resonators,
*Thomas Bunel, Matteo Conforti, and Arnaud Mussot*
10) Engineering Kerr Frequency Combs via Modal and Polarization Degrees of Freedom,
*Erwan Lucas and Julien Fatome*
11) Laser frequency combs for astronomy,
*Yuk Shan Cheng and Derryck T. Reid*
12) Microresonator-Filtered Lasers: Opportunities and Challenges,
*Alessia Pasquazi and Marco Peccianti*
13) Dissipative Solitons and Optical Frequency Combs in degenerate cavity VECSELs,
*M. Giudici, M. Marconi, A. Bartolo, N. Vigne, B. Chomet, A. Garnache, G. Beaudoin and I. Sagnes*
14) Developing Talent and Skills for Timing Technology Research and Implementation,
Richard Burguete and Sarah Hammer
15) Microcombs for Portable Optical Clocks, *Jonathan Silver*

# Towards Frequency Agile Solitons in Microresonators

Dmitry V. Skryabin[1,2,3]

[1] Department of Physics, University of Bath, Bath, BA2 7AY, UK
[2] Centre for Photonics, University of Bath, Bath, BA2 7AY, UK
[3] National Physical Laboratory, Teddington, TW11 0LW, UK

E-mail: d.v.skryabin@bath.ac.uk

**Status**

Optical solitons are a remarkable and widely observed phenomenon across various photonic platforms, enabled by the pulse-profile-dependent nonlinear shift of the refractive index, which generates pulse chirping that compensates for dispersive spreading. If a pump or gain is introduced, a second balance can be established, allowing for dissipation compensation. At the same time, the soliton is altered but preserved, and hence, it is now called a dissipative soliton [1]. Research into optical solitons has flourished since the past century's work on fibre communications and laser modelocking [2], and has continued into this century, where remarkable progress with low-loss photonic-integrated circuits (PICs) has provided scientists with opportunities to engineer dispersion and generate solitons at ever-lower energies in chip-scale devices [1]. Spectrally, modelocking of multimode signals into trains of short pulses is associated with optical frequency combs (OFCs), which consist of hundreds to thousands of equally spaced, narrow resonance lines used for timekeeping, precision spectroscopy, and metrology [3].

Over the past decade, OFC research has witnessed outstanding progress through the development of microresonators and the realisation that the microresonator OFC generation is explained by the existence of soliton solutions of the Lugiato-Lefever model, frequently referred to as dissipative Kerr solitons (DKSs) [3]. Microresonators are challenging fibre-laser OFC sources in their traditional application areas and also serve as a novel, compact, and energy-efficient multifrequency light source for optical information processing, see Fig. 1.

Spectral broadening of solitons and associated OFCs into an octave has been realised as an important challenge early on in the microresonator research and achieved through the emission of dual dispersive waves and small microresonator radii, increasing the comb-tooth separation to one THz [3]. However, spectral coverage remains one of the main limitations of DKSs since they are most commonly centred and extending around the telecom C-band window, where the geometry of the waveguides made using a range of thin-film materials can be engineered to be anomalous and thus be balanced by the positive Kerr nonlinearity for generating bright DKSs, see Fig. 1. Consequently, frequency tunability of solitons in microresonators, overcoming constraints imposed by anomalous dispersion, and enabling the generation of octave- and multi-octave OFCs—including in the normal dispersion regime—remain active areas of research, which we outline below limiting ourselves to passive microresonators pumped by a tabletop or chip-scale source.

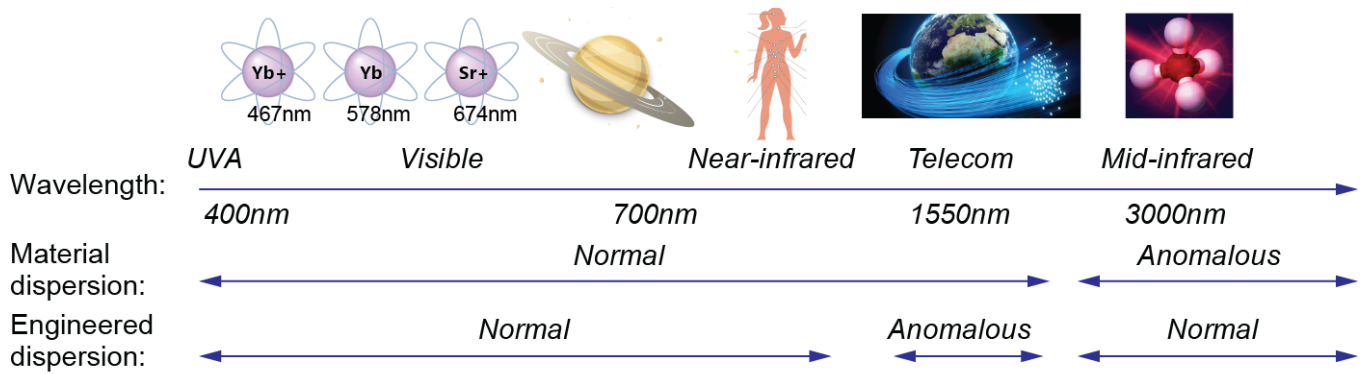

*Figure 1. Schematic illustration of some applications of OFCs and microresonator solitons vs wavelength and  of ranges of normal and anomalous material and engineered dispersions as in common silicon-nitride and lithium-niobate microresonators. Left to right: optical atomic clock transitions, astronomical observations, bio-imaging, optical telecommunications and molecular vibrations*.

**Current and future challenges**

Possible strategies to generate OFCs and solitons in new frequencies ranges outside telecom C-band are (i)  to use pump in the required range and engineer a suitable material platform and resonator geometry to comply with the soliton generation conditions and (ii) to use one of the frequency conversion processes to convert the pump spectrum and simultaneously generate OFCs and solitons centred at new frequencies. Regarding the first approach, we refer the reader to a roadmap chapter by Marris-Morini & Bousseksou on mid-IR OFC sources and mention predictions of waveguide geometries and mode crossings, which provide windows of anomalous dispersion for soliton generation in the visible range in sub-micron lithium-niobate (LN) waveguides [4,5]. As fabrication improves and expands into a broader range of materials, enabling sub-wavelength channel cross-sections for dispersion engineering and high-Q values from UVA to mid-IR, we should expect to see results on directly pumped combs across a diversity of wavelengths.

For the second approach, one can use four-wave mixing in $\chi^{(3)}$  materials, three-wave mixing in $\chi^{(2)}$ materials and the Raman effect. Either of these processes, along with their numerous special cases determined by the pumping and phase-matching conditions, allows for the generation of multicolour solitons and OFCs [6-18]. Examples and concepts of multicolour soliton generation in microresonators are schematically illustrated in Fig. 2. Dispersion engineering of waveguiding channels and periodic poling allow the establishment of desired phase-matching conditions, which can be optimised to a degree and mitigate the effects of group-velocity walk-off between different pulse components [19]. Early studies have reported soliton formation at 1550 nm, with the second harmonic comb being non-solitonic [8]. Multicolour solitons can take the form of bright-bright [6,11], dark-dark [6], and bright-dark [9] soliton pairs, as well as pairs of domain walls or switching waves [12]. More specifically, [6] used periodically poled LN microresonators with mixed normal and anomalous dispersion, demonstrating interlinked families of bright-bright and dark-dark soliton pairs at 1550 nm (pump) and 775 nm (second harmonic) and also two-colour breathers. Additionally, there have been reports of multicoloured modelocked waveforms in microresonators where the pump field remains quasi-continuous-wave while the desired signal consists of soliton pulses [13-15]. E.g., [13] reported half-harmonic generation in AlN (780 nm pump, 1560 nm signal), and the soliton formation was observed in the signal while the pump remained non-solitonic, where both frequencies fall into the anomalous dispersion range. Various aspects of multicolour soliton effects reported in [6,15]  highlight that a relatively short resonator length, which is not significantly larger than the pulse length, can profoundly alter the soliton formation mechanisms if compared to textbook cases assuming solitons in infinite geometries [20].

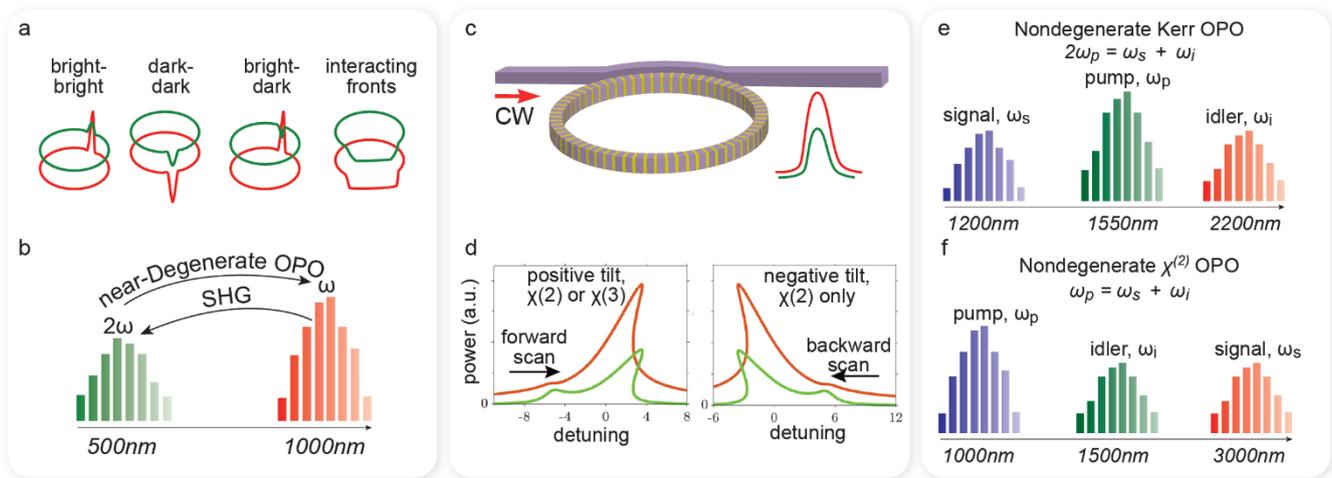

*Figure 2 **a, b.** Illustration of two-colour solitons in ring geometries associated with second-harmonic generation (SHG) and degenerate parametric down-conversion. **c, d.** Illustration of soliton generation in a χ(2) microresonator with periodic poling and phase-matching–controlled positive or negative nonlinear shifts of the resonance lines. **e, f.** Generation of multi-colour optical frequency combs (OFCs) in non-degenerate parametric oscillators in Kerr and χ(2) microresonators.*

**Advances in science and technology to meet challenges**

Knowledge about dispersion and loss engineering on the short-wavelength side of the spectrum, as well as the achievable phase-matching conditions for three- and four-wave mixing processes, remains sparse. Fabrication of samples in new geometries designed for UV and visible OFCs and related loss improvement work is ongoing. For example, current loss achieved in LN waveguides in the C-band is <10dB/cm, and it can be reduced further (material absorption of 0.5dB/m stays nearly flat to 400nm and starts sharply rising after), giving a theoretical limit for the resonator quality factors (Q) above 100 million. The best practical demonstration of the intrinsic Q in the LN resonator was 29 million [21]. However, intrinsic Q values around and below one million are already sufficient for OFC generation and are realistic to achieve in the visible. Establishing that periodic poling and temperature tuning of resonance in a *χ(2)* microresonator is capable delivering either positive or negative sign of the phase matching parameter and hence providing an on-demand red or blue shift of the resonance frequencies is a conceptual step in the OFC area [6] marking a way to depart from the constraints of the Kerr effect capable of only red shifts, see Figs. 2c,d. Further research is required to demonstrate the practical potential of this technique for soliton generation in the normal dispersion range and under conditions of varying dispersion signs across the OFC spectrum. The full strength and advantages of *χ(2)* nonlinearity show only at the phase-matching conditions. Therefore, engineering poling profiles and choosing materials for strong broadband nonlinear response needs further studies and demonstrations.

Parametric frequency conversion in both *χ(2)* and *χ(3)* cases can be degenerate and non-degenerate with the non-degenerate case providing generation of the broadly separated signal and idler fields which spectral location is determined by the phase matching conditions, see Figs. 2e,f. Half- and second-harmonic generation are typical examples of the degenerate processes, see Fig. 2b. Tunability of the degenerated processes is always restricted by the tunability of the pump sources and phase-matching only assists with conversion efficiency. Contrary, phase-matching conditions for non-degenerate parametric conversion can be tuned with geometry and temperature and thus provide a way to control where signal and idler frequency are generated while the pump can be kept fixed. This has been actively explored in the continuous wave regimes of operation of both χ(2) and χ(3) microresonator parametric oscillators, see, e.g., references in [15]. Experimental studies of multicolour solitons in non-degenerate Kerr case has just been initiated [15], while solitons in nondegenerate χ(2) parametric processes in microresonators are yet to be investigated. Achieving multicolour OFCs via third harmonic generation and three-photon parametric down conversion also represents significant challenge and interest. Thus there are many opportunities for theoretical and experimental research on nonlinear frequency conversion scenarios in microresonators.

**Concluding remarks**

The work on the development of any-wavelength microresonator OFC sources is gaining momentum (see also the roadmap chapters by Marris-Morini & Bousseksou and Bres), with single and multi-component solitons at the core of these studies. Exploiting χ(2) nonlinearity for these purposes is also rapidly advancing. Lithium niobate currently appears as a preferred thin-film χ(2) material with other closely following and competing platforms, such as, e.g., aluminium nitride, gallium arsenide, and lithium tantalate [22]. These developments are occurring with an emerging understanding that microresonator OFCs are challenging the traditional understanding of optical solitons, as their pronounced, tens of GHz to one THz, spectral discreteness allows for the existence of modelocked pulses in forms and under conditions which are more forgiving than those assumed for solitons with continuous or quasi-continuous spectral content.

**Acknowledgements**

We acknowledge support from the Royal Society (Grant No. IES/R3/223225) and UKRI EPSRC (Grant No. EP/X040844/1).

# Optical Microcombs for Precision Metrology Applications

Arne Kordts[1], Richard Zeltner[1], Ronald Holzwarth[1]

[1] Menlo Systems GmbH, Martinsried, Germany

E-mail: a.kordts@menlosystems.com

## Status

Optical frequency combs have revolutionized precision metrology and spectroscopy, enabling unprecedented advancements in optical clocks and serving as a crucial enabling technology for quantum computing platforms [1,2].

To date, the highest-performing comb systems for these demanding applications have typically relied on femtosecond mode-locked lasers (MLLs), where pulse formation is dominated by passive mode locking. However, emerging microcomb platforms — pulsed laser systems based on microscale resonators — offer compelling opportunities, particularly when implemented on photonic integrated circuit (PIC) platforms. Such microcombs promise to realize high-precision comb technologies in exceptionally compact form factors that can be fabricated at scale [3]. This development opens new frontiers for widely deployed applications, including compact and precise optical clocks suitable for navigation systems such as the Global Positioning System (GPS), as well as ultra-stable photonically generated microwave signals for radar systems.

For these high-precision applications, it is unlikely that any pulsed laser source alone will inherently fulfill the stringent noise performance requirements. Instead, lasers achieve the necessary stability by locking to external optical frequency references, thereby inheriting their exceptional stability characteristics. Typically, this involves linking to ultra-stable optical cavities for short-term stability, narrow optical atomic transitions for long-term stability, or a combination of both. Additionally, self-referencing techniques are employed to transfer this stability effectively across the entire comb spectrum. Consequently, high-precision comb sources are not standalone components, but rather sophisticated systems comprising specialized sub-modules.

To effectively link to external references and transfer their stability, the comb source itself must possess inherently low passive noise and must provide rapid, active control in its two fundamental degrees of freedom: the carrier-envelope offset (CEO) frequency and the repetition rate. Microcomb platforms targeting high-precision applications will need to meet the same criteria. In particular, among the various PIC-based microcomb systems [4], only those whose pulse formation dynamics are primarily governed by soliton generation [5] are considered promising candidates for these demanding tasks.

Initial research efforts have primarily focused on reliably initiating and maintaining soliton states, as well as generating these states with high power efficiency. Consequently, robustly transferring the stability of state-of-the-art optical references onto microcomb platforms remains an ongoing challenge. Addressing this frontier will necessitate implementing dedicated actuators capable of independently and rapidly controlling both fundamental degrees of freedom of the comb spectrum.

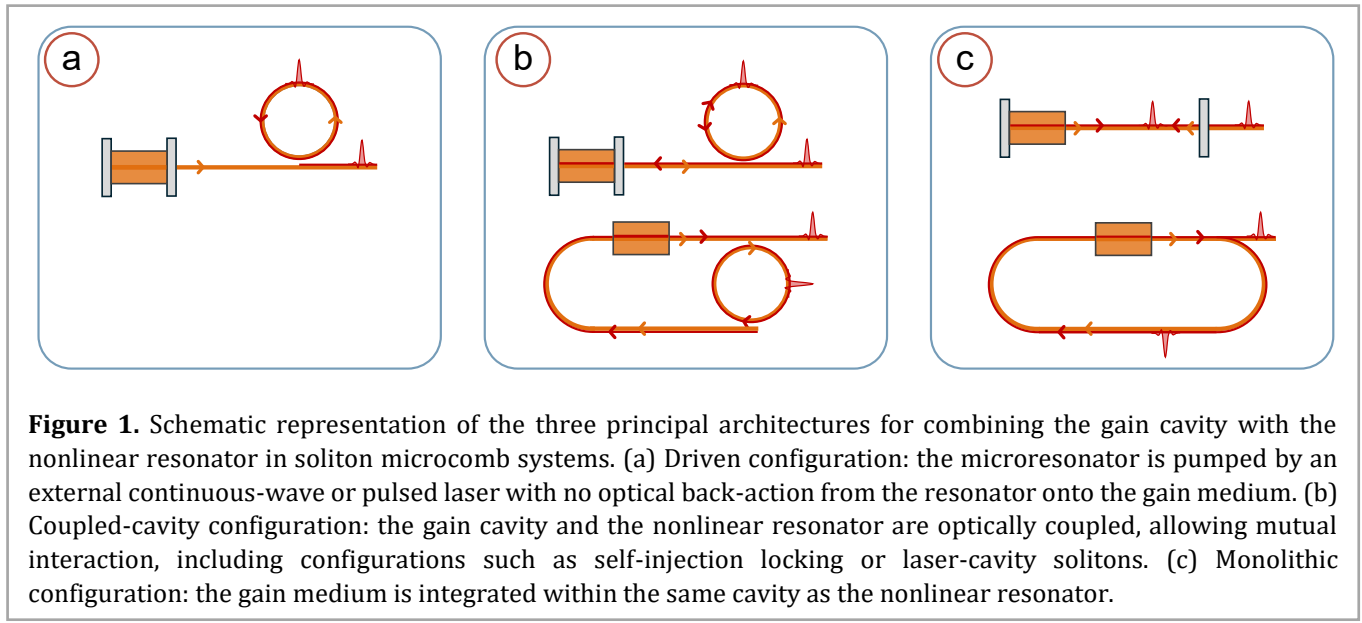

**Figure 1.** Schematic representation of the three principal architectures for combining the gain cavity with the nonlinear resonator in soliton microcomb systems. (a) Driven configuration: the microresonator is pumped by an external continuous-wave or pulsed laser with no optical back-action from the resonator onto the gain medium. (b) Coupled-cavity configuration: the gain cavity and the nonlinear resonator are optically coupled, allowing mutual interaction, including configurations such as self-injection locking or laser-cavity solitons. (c) Monolithic configuration: the gain medium is integrated within the same cavity as the nonlinear resonator.

**Current and future challenges**

It is beneficial to classify soliton based microcomb architectures according to how the optical gain is integrated with the nonlinear microresonator as shown in Figure 1. We distinguish three primary architectures: (1) the driven configuration, (2) the coupled configuration, and (3) the monolithic configuration.

In the driven configuration, the microresonator dynamics are isolated from the gain medium, meaning there is no direct optical back-action from the resonator onto the pump laser source. This is currently the most established approach, typically employing continuous-wave (CW) pumping [5], pulsed pumping [6], or more recently, advanced multi-color pumping schemes. Progress over the past decade has led to robust and repeatable soliton generation, enabling a shift in focus toward achieving fine control over the comb spectrum. A key advantage of this architecture lies in its natural compatibility with external frequency references. Since the comb spectrum is coherently generated from the pump laser, stabilizing the pump laser frequency directly stabilizes one degree of freedom of the comb. This principle can be extended through repetition-rate injection locking, thus linking the second degree of freedom to an external microwave reference. More recently, Kerr-induced synchronization (KIS)—enabled by multi-color pumping—has demonstrated the ability to control the complete soliton spectrum via two pump lasers [7,8]. This approach has enabled a direct path for full stabilization and self-referencing of the comb [9], although current implementations remain limited by the noise of pump lasers that are not yet stabilized to ultra-stable optical cavities.

The coupled configuration, exemplified by self-injection locking [10,11] and laser cavity-soliton microcombs [12], involves direct optical interaction between the gain cavity of the pump laser and the nonlinear microresonator. This configuration is inherently power efficient, requiring no additional engineering to suppress unwanted pump radiation, and typically enables simplified initiation and maintenance of soliton states. However, a current shortcoming is the absence of dedicated actuators for fast, independent control of the comb's degrees of freedom [13]. Addressing this challenge will be crucial for unlocking the coupled configuration's full potential for precision metrology.

In the monolithic configuration, the gain medium is integrated directly within the nonlinear cavity that generates the pulse train. This architecture is well established in traditional mode-locked lasers, where intra-cavity modulation schemes for full comb-spectrum control are well established.

Further, the most mature PIC-based mode-locked lasers — those based on III-V semiconductor platforms [4] — also follow a monolithic design. However, in these III-V semiconductor MLLs, pulse formation is primarily governed by the nonlinear response of semiconductor-based saturable absorbers (SAs). The intrinsic noise properties of these SA are a limiting factor, making these systems unsuitable for high-precision applications.

Nevertheless, significant noise performance improvements have recently been achieved through hybrid integration with low-loss silicon or silicon nitride platforms [4]. These advances enhance the role of Kerr nonlinearity in the pulse formation process [14], pointing to a potential path away from reliance on semiconductor SAs and toward soliton operation. In parallel, efforts have been made to directly replicate the architecture of fiber-based MLLs on-chip, aiming to combine integrated rare-earth-ion-based gain media with passive mode-locking elements based on nonlinear interferometers [15]. However, the realization of stable soliton operation in such monolithic PIC systems remains an open challenge. As a result, the implementation of established actuator schemes for full-spectrum control has yet to be demonstrated in this context.

**Advances in science and technology to meet challenges**

While soliton-based microcombs have demonstrated major progress in recent years, realizing their full potential for high-precision applications still requires addressing configuration-specific integration and control challenges. In the driven configuration, the key challenge is to consolidate the various techniques developed individually into a fully integrated, coherent system.

This involves combining low-noise lasers with fast modulation capability, integrating them with a low-noise microresonator, and ensuring efficient use of pump power [16,17]. Moreover, the resonator must be stabilized—either passively through thermal and mechanical design or actively via feedback—to maintain the soliton state over time. A further requirement is the implementation of deterministic and reliable soliton initiation procedures. Only once these building blocks are combined in a robust and scalable platform will the driven architecture fulfill their promise as compact and stable sources for field-deployable precision systems.

The coupled configuration, though inherently power-efficient and well suited for soliton generation, has yet to demonstrate the capability for full and high-bandwidth control of the comb spectrum. Meeting this challenge will likely require the introduction of intracavity modulation — either in the laser cavity or the microresonator. Direct integration of modulation structures into the low-loss passive resonator is technologically challenging and may not be feasible within current fabrication platforms. However, the potential use of Kerr-induced synchronization in this architecture could offer an alternative means for spectral control. Another promising path lies in integrating modulators into the laser cavity itself. Drawing from concepts established in fiber-based MLLs, it may be possible to achieve effective control by selectively modulating the group velocity without perturbing the phase velocity. Demonstrating such a capability in coupled microcomb systems would mark an important step toward their use in metrological applications.

The monolithic configuration presents the most demanding challenge among the three. In this architecture, all essential functional components—low-noise gain, nonlinear propagation, and fast modulation—are co-integrated within a single cavity. Their tight interdependence makes modular optimization extremely difficult and places stringent requirements on material compatibility and fabrication precision. For monolithic designs to become viable for high-precision applications, substantial advancements in material platforms are required. Platforms that simultaneously support strong $\chi^{(3)}$ nonlinearity for soliton formation and $\chi^{(2)}$ or electro-optic effects for fast modulation are especially promising [18]. Realizing this level of integration would open a path to compact, high-performance microcombs fully analogous to traditional mode-locked fiber lasers, but in a scalable on-chip format.

**Concluding remarks**

The recent progress across various solitonic microcomb platforms paints an optimistic picture for their future role in high-precision applications. Advances in soliton generation, spectral control, and integration—combined with growing engagement from the research community—underscore the momentum in this field. Several of the requirements for precision-grade microcombs have been demonstrated individually, including turnkey soliton operation, high power efficiency, and fast spectral control compatible with linking to optical references. Among the explored architectures, the driven and coupled configurations appear to be the most viable candidates for near-term deployment in high-precision systems, as they strike a favorable balance between performance and complexity. In contrast, the monolithic configuration remains less mature and is unlikely to see immediate use in demanding applications due to the significant co-integration challenges it presents—particularly in combining low-noise gain, nonlinear media, and fast modulators within a single cavity. Nonetheless, with continued progress in material platforms and fabrication techniques, monolithic designs may eventually re-emerge as a compelling option for fully integrated microcomb systems.

**Acknowledgements**

The authors acknowledge financial support by the Federal Ministry of Research, Technology and Space of Germany (BMFTR) in the project LICHTBRIQ and by the European Union through the Eurostars programme with contract number E!113569 (COSMIC).

# Ultimate Coherence Performance Limits of Soliton Microcombs

Victor Torres-Company[1], Tobias Herr[2], Fuchuan Lei[3] and Qi-Fan Yang[4]

[1] Department of Microtechnology and Nanoscience, Chalmers University of Technology, SE-41296 Gothenburg, Sweden
[2] Deutsches Elektronen-Synchrotron DESY, Notkestr. 85, 22607 Hamburg, Germany and Department of Physics, Universität Hamburg, Luruper Chaussee 149, 22607 Hamburg, Germany
[3] State Key Laboratory of Integrated Optoelectronics and School of Physics, Northeast Normal University, Changchun 130012, China
[4] State Key Laboratory for Artificial Microstructure and Mesoscopic Physics and Frontiers Science Center for Nano-optoelectronics, School of Physics, Peking University, Beijing 100871, China

E-mail: torresv@chalmers.se

**Status**

Microcombs promise to attain, on a chip-scale platform, the same level of frequency instability and accuracy as laser frequency combs produced using traditional passively mode-locked fiber or solid-state lasers. By harnessing the physics of dissipative Kerr solitons in high-Q microresonators, they have emerged as strong contenders for the realization of next-generation systems-on-chip, with applications spanning optical communications, microwave photonics, optical clocks, and quantum technologies.

Over the past decade, the community has developed a deeper understanding of the fundamental noise sources that limit the frequency stability and coherence of these comb sources. Thermorefractive noise, quantum shot noise, and the phase and intensity noise of the pump laser impose strict limits on optical linewidth and timing jitter [1,2]. These noise mechanisms are intrinsic to microcomb generation and set bounds on the purity of the generated microwave signals and can affect the system performance when they are implemented in optical clocks.

In free-running microcombs (i.e., without stabilization), these noise sources contribute to an increasing optical phase noise with mode number [2] — akin to the elastic tape model originally formulated for mode-locked lasers. As a result, the timing jitter degrades the repetition rate stability of the pulse train and hence the phase noise of microwave carrier generated by photodetection. To address this, an operation regime known as *quiet point* has been devised [3], where the sensitivity of the comb to detuning and pump noise is minimized. These points arise from the backaction of dispersive wave emission, which imparts a recoil on the soliton and creates a stabilizing negative feedback loop. This effect suppresses both the transduction of pump noise and quantum fluctuations into repetition rate jitter [4]. Additional strategies such as laser cooling [5] have been explored to mitigate the effects of thermorefractive noise on the microcomb stability.

Despite these advances, the frequency stability of state-of-the-art free-running microcombs still falls short compared to that of conventional mode-locked laser combs and microwave signals derived from dielectric resonator oscillators. This performance gap has recently motivated efforts to integrate active stabilization schemes based on a process known as two-point frequency division [6,7], Fig. 1. In this approach, two optical modes of the comb are locked to external references, such as narrow-linewidth lasers or stable cavities, enabling a tight phase locking across the entire comb. This scheme can be elegantly implemented based on nonlinear synchronisation via sideband-injection of a second pump laser, enabling performance below the thermorefractive noise limit of the microresonator [8-10]. This strategy offers a path to significantly reduce phase noise and drift, and it is increasingly seen as essential to push microcombs beyond their (free-running) performance limits.

**Current and future challenges**

Microcombs offer the possibility of optical-to-microwave frequency division, where the phase noise of the generated microwave signal becomes a scaled-down replica of the optical references. In two-point locking schemes, this division factor is approximately set by the number of comb lines. For microcombs spanning the C-band with a 20 GHz repetition rate, this implies a theoretical phase noise reduction of ~46 dB. Recent demonstrations exploit this property by locking two comb lines to a common optical reference [11,12], which transfers its long-term stability to the microwave domain, while short-term noise becomes limited by photodetection shot noise. As of now, it is not clear whether combs operated in the anomalous or normal dispersion regime will ultimately enable lower timing jitter, and to what extent unavoidable imperfections in the resonator will limit the noise performance.

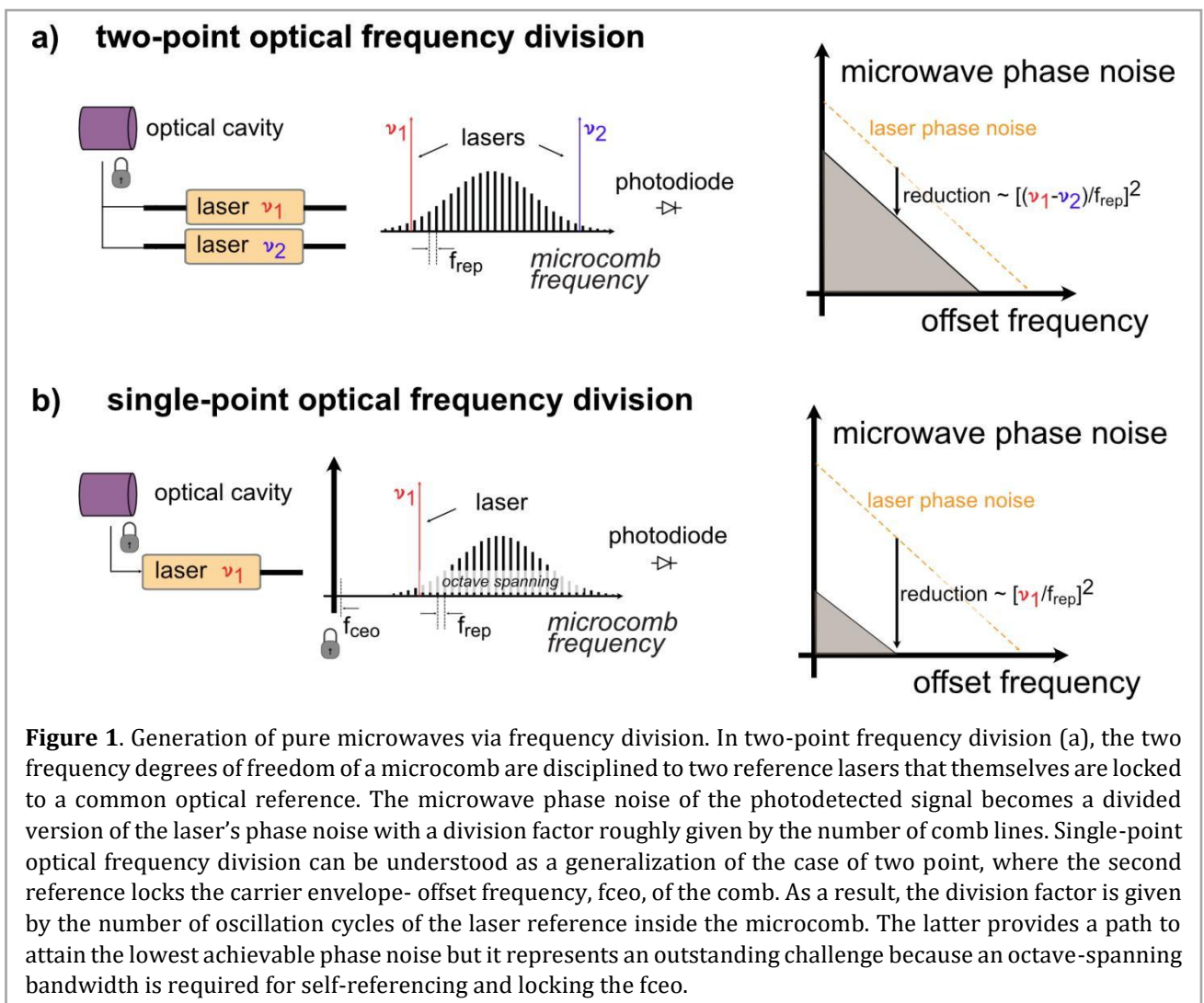

**Figure 1**. Generation of pure microwaves via frequency division. In two-point frequency division (a), the two frequency degrees of freedom of a microcomb are disciplined to two reference lasers that themselves are locked to a common optical reference. The microwave phase noise of the photodetected signal becomes a divided version of the laser's phase noise with a division factor roughly given by the number of comb lines. Single-point optical frequency division can be understood as a generalization of the case of two point, where the second reference locks the carrier envelope- offset frequency, fceo, of the comb. As a result, the division factor is given by the number of oscillation cycles of the laser reference inside the microcomb. The latter provides a path to attain the lowest achievable phase noise but it represents an outstanding challenge because an octave-spanning bandwidth is required for self-referencing and locking the fceo.

Pushing the noise floor even lower demands advances in multiple areas. Thermorefractive noise of the reference cavity remains a fundamental limit [6], and mitigating it requires ultra-low-loss (high-Q) cavities with large mode volumes. Similarly, high-power, high-linearity photodiodes with broad bandwidth are needed to faithfully convert optical pulses into low-noise electrical signals. Importantly, all these components must be compatible with heterogeneous integration to enable scalable and manufacturable solutions.

Microcombs differ critically from traditional passively mode-locked lasers in that their repetition rates are much higher. This presents unique metrology challenges. First, the pulse-to-pulse fluctuations (timing jitter) manifest at offset frequencies well beyond the range of conventional phase noise analyzers. Second, the much lower pulse energies reduce the sensitivity of standard nonlinear techniques used to characterize timing noise. Finally, a proper assessment of purity now requires combining several diagnostic methods across wide frequency spans because the need to integrate phase noise from $\sim 1$ Hz all the way up to the Nyquist frequency.

Ultra-low-noise performance in conventional frequency combs is obtained via single-point optical division, where one optical mode is locked to a reference and the entire comb inherits its stability; this effectively maximizes the frequency interval that is utilized for optical division. Realizing this in microcombs, however, requires octave-spanning spectra and self-referencing at high repetition rates. If achieved, the division factor could exceed 80 dB for C-band references and 20 GHz combs, enabling the purest microwave signals ever demonstrated on-chip at room temperature.

Attaining this performance level will likely challenge our fundamental understanding of the quantum nature of optical oscillators at the chip scale. Microcombs exhibit deep quantum correlations between modes [13] that defy mean-field descriptions such as the Lugiato-Lefever equation. Bridging this gap between quantum optics and classical modeling represents both a theoretical and experimental frontier.

**Advances in science and technology to meet challenges**

Recent progress in silicon nitride waveguide technology has enabled propagation losses below 1 dB/m in meter-long waveguides [7,14], paving the way for ultra-high-Q cavities with large mode volumes. However, the thin-core geometries used to minimize loss are typically incompatible with the dispersion engineering needed for octave-spanning spectra, which are essential for self-referencing.

To overcome this trade-off, multilayer photonic integration is emerging as a powerful solution [15,16]. By stacking waveguide layers of different thicknesses at the wafer level, this approach allows independent optimization of loss and dispersion [15]. Furthermore, three-dimensional integration techniques make it possible to combine disparate material platforms (including III-V semiconductors [16] and ferroelectrics [17]) with silicon nitride in a monolithic process. This opens new possibilities for co-integrating modified uni-traveling carrier photodiodes directly with low-loss waveguides [18]. These photodiodes offer the rare combination of high bandwidth, high responsivity, and high power-handling capability without significant nonlinear distortion, which are ideal characteristics for detecting ultrafast optical pulse trains with minimal noise degradation.

Another area of active development involves coupled-cavity designs [19] and dual combs [20]. These innovations can overcome the power scaling limitations of conventional soliton microcombs and facilitate the generation of octave-spanning spectra at photodetectable repetition rates. By increasing the line density and energy conversion efficiency, they bring practical self-referencing within reach for chip-scale devices.

On the theoretical side, new insights into the quantum properties of microcombs are beginning to reshape our understanding of their noise performance. Studies of intermodal quantum correlations [14], including squeezing and entanglement, suggest that microcombs may operate in regimes not well described by classical mean-field models such as the Lugiato-Lefever equation. These findings point to the need for a quantum optics framework to accurately predict the ultimate limits of coherence and stability.

Together, these advances in materials, device integration, architecture, and theory are laying the foundation for a new generation of microcombs with performance approaching (or potentially exceeding) that of bulk optical systems, but in a scalable, chip-integrated format.

**Concluding remarks**

Microcombs have evolved into a leading platform for compact, scalable frequency comb generation. Their potential to deliver ultralow-noise optical and microwave signals on chip makes them uniquely positioned for emerging applications in timekeeping, navigation and communications.

Despite significant progress in understanding and mitigating noise sources, the performance of microcombs still lags behind that of bulk (non-integrated) modelocked fiber or solid-state lasers. Closing this gap requires advances in integrated photonics, including high-power handling, high bandwidth photodetectors and ultra-high-Q resonators.

Looking ahead, single-point optical frequency division promises to unlock new regimes of performance, but further innovations in heterogeneous are required. As these capabilities mature, microcombs could redefine the practical limits of frequency division and signal purity, enabling chip-scale systems with coherence levels once reserved for laboratory-scale instruments.

**Acknowledgements**

This work has been supported from the Swedish Research Council (2020-00453, 2022-06575), the Air Force Office of Scientific Research (FA8655-23-1-7012), and the Swedish Foundation for Strategic Research (IS24-0075). T.H. acknowledges funding by the Helmholtz YIG program (VH-NG-1404). FL acknowledges funding from the National Natural Science Foundation of China (NSFC) (62475042) and the 111 project (B25030).

# Kerr Combs in Microresonators with Linearly Coupled Modes

Tobias Herr[1,2], Erwan Lucas[3], Victor Torres Company[4]

[1] Department of Physics, University of Hamburg, Hamburg, Germany
[2] Deutsches Elektronen-Synchrotron DESY, Hamburg, Germany
[3] Laboratoire Interdisciplinaire Carnot de Bourgogne ICB UMR 6303, Université Bourgogne Europe, CNRS, F-21000 Dijon, France
[4] Department of Microtechnology and Nanoscience, Chalmers University of Technology, SE-41296 Gothenburg, Sweden

E-mail: tobias.herr@desy.de, erwan.lucas@u-bourgogne.fr, torresv@chalmers.se

**Status**

The formation of dissipative temporal patterns in microresonators enables the generation of optical frequency combs. These 'Kerr combs' or 'microcombs' can achieve pulse repetition rates of over 10 GHz and, thanks to the parametric nonlinear generation process, offer extremely low noise levels. Their potential for integration into chips as waveguide-based resonators makes them a key element in the widespread application of advanced photonic technology. Microcomb generation can be viewed as a phase-coherent, nonlinear energy exchange and synchronisation between many optical modes, at least one of which is driven by an external, usually continuous-wave (CW) pump laser. The dispersion of the resonator plays a decisive role in these nonlinear processes as it impacts the phase-matching between different modes and therefore the efficiency and phase of the nonlinear conversion processes.

Previous work has leveraged dispersion engineering based on waveguide cross-section and bend radius, leading to spectra spanning an octave and operation across a wide range of central wavelengths. Recently, the toolkit for designing resonator dispersion has been expanded by utilising linear coupling between resonator modes (Fig. 1). This includes coupling between counter-propagating modes in photonic crystal ring resonators [1], deliberate coupling between fundamental and higher-order modes [2], and coupling between distinct resonators [3]. In all cases, hybridisation of the modes occurs, leading to a new set of eigenmodes (supermodes) with modified frequencies, and thus a modified resulting dispersion.

Unlike traditional approaches that globally tune dispersion via waveguide geometry or bend radius, mode hybridisation enables spectrally localised and potentially mode-by-mode control [4,5]. This capability has been used for spectral shaping of the comb envelope [5–7], creating highly efficient systems [8], suppressing undesired nonlinear states [1], achieving deterministic self-injection locking of combs [9], generating single-mode dispersive waves [10], reducing the comb noise [11] and generating Kerr combs in the normal dispersion regime via switching waves [2,3,12]. Furthermore, novel phenomena such as soliton hopping between resonators [13] and topological combs [14] depend on linear coupling between resonator systems. In summary, designing the linear coupling between modes influences the nonlinear coupling that underpins comb generation, and in some cases, the coupling to the external laser, adding new functionalities to Kerr combs. Engineering the linear mode coupling in microresonators ultimately promise to yield chip-integrated systems that can reliably and efficiently produce tailored, broadband comb spectra with low noise across the material transparency window.

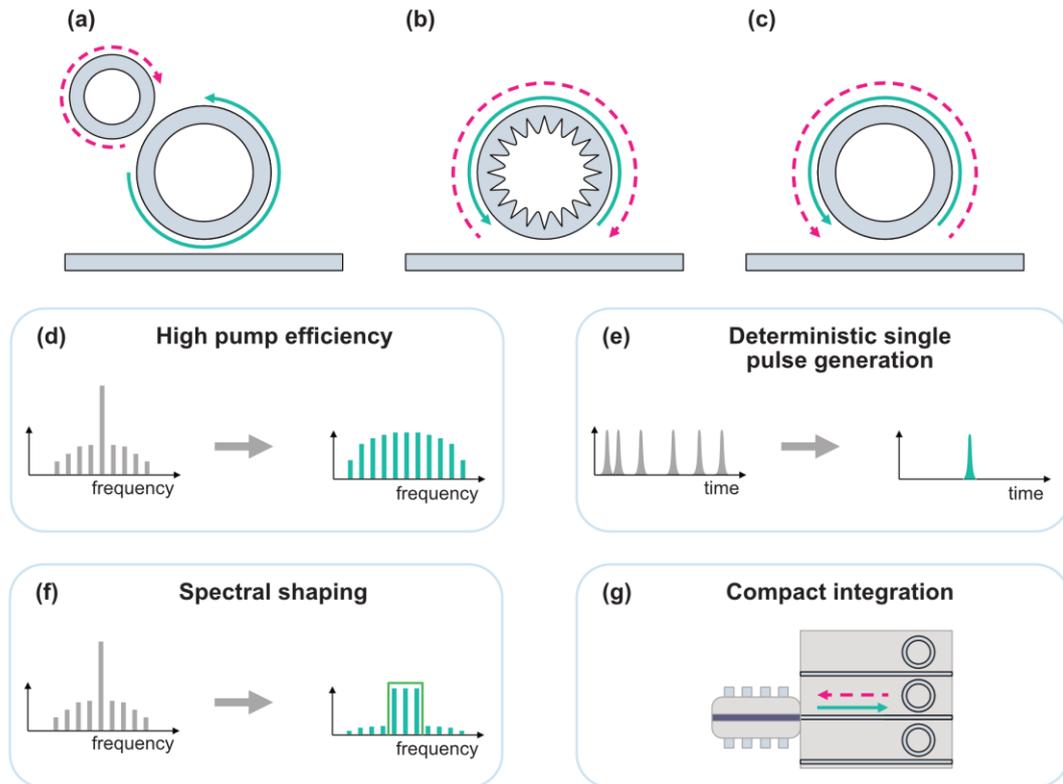

**Figure 1. Kerr-combs in coupled microresonators.** Examples of coupled microresonator systems include **(a)** Coupled microresonators, **(b)** spectrally-selective coupling of forward and backward propagating modes in a photonic crystal ring microresonator, and **(c)** coupling between higher-order modes propagating in the same direction. Coupled microresonator systems have enabled **(d)** high pump efficiency, **(e)** deterministic single pulse generation, **(f)** spectral shaping, and **(g)** compact chip-integrated systems.

**Current and future challenges**

Kerr-combs are usually modelled in the framework of the Lugiato-Lefever equation (LLE), a damped, driven nonlinear Schrödinger equation. In its original form, the LLE includes group velocity dispersion (second-order dispersion) as the first relevant dispersion term. The dynamics and solutions of the LLE are well understood, including its bifurcation characteristics and phase diagram as a function of the key operational parameters: pump laser detuning and pump laser power. When higher-order dispersion terms are considered, the nonlinear dynamics of the resonator is modified, leading, for instance, to the formation of dispersive waves. The solutions, however, still resemble those of the original LLE.

In contrast, the strong local nature of the dispersion modification induced by mode hybridisation can substantially transform the solution space and the system's phase diagram. While dynamic models can readily include such effects, a detailed understanding of the structure and existence of stable solutions, as is the case for the LLE, is currently lacking.

Mode hybridisation not only leads to a new set of hybridized modes or supermodes with modified dispersion, but also introduces new energy exchange processes between the coupled modes resulting in non-trivial nonlinear dynamics. For example, during comb initiation in photonic crystal resonators, where a hybridised pump mode provides access to the desired single-pulse states, complex energy-exchange dynamics have been observed between forward- and backward-propagating modes [15]. The power injected in the system is spread across the coupled modes, which typically increases the threshold needed to trigger the initial comb formation [9]. Linear coupling of a mode in the wing of a Kerr comb to higher-order modes can create undesirable breathing dynamics [16]. A better understanding of such processes is needed to realise the potential of linear mode coupling. Additional challenges arise from differential mode frequency shifts, for instance, when coupling to higher-order modes with different thermo-optic coefficients, impacting the system's noise dynamics [17,18]. Finally, in self-injection locked system [19] as well as microresonators nested in gain-loops [20], the linear coupling extends to an additional laser gain cavity, adding another dimension and laser gain dynamics to the system.

From an experimental perspective, new procedures and diagnostic tools are required to operate and characterise the emerging class of novel nonlinear systems. This includes, in particular, exploring the possible nonlinear states and reliably initiating them. For example, maintaining control of detuning in coupled resonators and managing parasitic linear couplings, such as those originating from reflection at the chip's facet, will be important. Finally, as mode coupling relies on advanced nanofabrication (i.e. coupling gaps between resonators and subwavelength photonic crystal structures), stable and reproducible photonic chip fabrication is essential for this technology to flourish.

**Advances in science and technology to meet challenges**

So far, the ability to accurately model nonlinear dynamics using relatively simple models has been instrumental in advancing the field through a fruitful interplay between experimentation and numerical modelling. Although coupled systems are more complex, existing dynamic simulation frameworks are powerful tools that provide insight into the complex dynamics of coupled systems, and these tools can be extended. Furthermore, advanced stationary solvers and bifurcation analysis can play a pivotal role in exploring the multidimensional parameter space of coupled systems. The solid understanding of the LLE and the methods developed in this context provides a good starting point. Identifying an appropriate (super-) mode-basis in the presence of both linear and nonlinear coupling will be crucial in establishing models that are both efficient and intuitive. As laser gain media are incorporated into systems, the differences between Kerr-combs and CW or pulsed lasers become less distinct, indicating possible fruitful connections with fields such as semiconductor and fiber lasers.

Accurate sample characterisation, especially of the (hybridised) mode structure, is essential to establish a close link between minimal complexity viable models and the experiments. Additionally, reliable, reproducible and scalable sample fabrication with minimal defects inducing uncontrolled mode coupling is required. Recent work has already demonstrated the potential of wafer-level and foundry-based ultraviolet stepper lithography to address this requirement, at least within the telecommunications wavelength band and on the prevalent silicon nitride platform. Further advances towards smaller critical dimensions, initially accessible via electron beam lithography, have the potential to extend microresonators with linearly coupled modes towards shorter visible and ultraviolet wavelengths. New material platforms, such as lithium niobate and lithium tantalate, which are emerging alongside an advancing understanding of coupled systems, can further extend the scope of opportunities by offering second-order nonlinear effects too.

**Concluding remarks**

Microcombs based on single microresonators (or, more precisely, a single transverse mode family in a microresonator) have proven to be a prolific area of research and a catalyst for emerging applications. The additional parameter space in microresonator systems with linearly coupled modes now provides highly customisable access to nonlinear dynamics that go beyond the scope of the LLE, representing a fascinating research field at the interface of nonlinear system science, nanofabrication, and nonlinear photonics.

Supermode-engineering creates unprecedented technological opportunities for Kerr combs to combine chip-level integration, scalability and efficiency with enhanced low-noise performance and deterministic operation. Such properties are critical for applications including for instance in future communication networks and data centres, as well as for enabling technologies in science, such as the synchronisation of large-scale accelerators fascilities. Moreover, advances in Kerr-combs with linearly coupled modes may benefit from and perhaps also inspire other research domains such as, semiconductor laser and chip-integrated mode-locked lasers.

**Acknowledgements**

T.H. acknowledges funding from the European Research Council (ERC, grant No 853564), and the Helmholtz YIG program (VH-NG-1404). E.L. acknowledges the ANR – FRANCE (French National Research Agency) for its financial support of the SMARTKOMBS project ANR-23-CE24-0005, and funding from the European Union's Horizon Europe research and innovation programme under grant agreement No 101137000. V.T.C. acknowledges support from the Swedish Research Council (2020-00453, 2022-06575).

# Integrated Resonant Up-Conversion of Kerr Frequency Combs

**Camille-Sophie Brès[1]**

Institute of Electrical and Micro Engineering, Ecole Polytechnique Fédérale de Lausanne, Lausanne, Switzerland

E-mail: camille.bres@epfl.ch

**Status**

The fundamental demonstration of Kerr comb generation in microresonators in early 2000s [1], followed by the transition to chip-scale microresonators promised compact and robust frequency comb sources [2]. Nowadays such sources, battery powered, are routinely demonstrated [3]. Still most devices are confined to near-infrared wavelengths where silicon photonics excels, and Kerr comb operating in the middle infrared or visible ranges are scarce. While octave spanning frequency combs can be obtained, they have large free spectral range (~THz) and limited power per line. Applications in atomic spectroscopy, quantum optics, and biomedical imaging would highly benefit from efficient visible combs. Direct generation of Kerr combs at short wavelengths (Figure 1a) however faces fundamental challenges including increased material losses and predominantly normal dispersion that inhibit soliton formation. Quadratic frequency combs can also be directly generated through a combination of second-harmonic/sum-frequency generation (SHG/SFG) and optical parametric oscillation (OPO) (Figure 1b), benefiting from relatively low power threshold and higher pump-to-comb efficiency [4]. They however face complex phase-matching requirements and short-wavelength pumping needs. An alternative approach is to up-convert telecom Kerr combs to shorter wavelengths while preserving their properties (Figure 1c).

$\chi^{(3)}$ processes such as third-harmonic generation can convert combs and be implemented in any materials, but suffer from low efficiency. The main approach is hence to leverage $\chi^{(2)}$ nonlinear processes, primarily SHG, which exhibits higher conversion efficiencies. It however requires materials with broken inversion symmetry and careful phase-matching considerations. Still simultaneous generation and doubling of Kerr frequency combs within ring resonators has been demonstrated across multiple integrated photonic platforms. Miller et al. pioneered this concept in silicon nitride microresonators by utilizing surface $\chi^{(2)}$ to achieve simultaneous comb generation and doubling, though with limited conversion efficiency due to the weak surface nonlinearity [5]. He et al. demonstrated significant advancement using thin-film lithium niobate microresonators, leveraging the material's strong $\chi^{(2)}$ to achieve efficient harmonic conversion while maintaining soliton comb operation [6]. This approach was extended to gallium phosphide, exploiting the material's wide transparency window and strong $\chi^{(2)}$ [7], and to aluminum nitride microresonators [8].

Further advances in integrated resonant upconversion of broadband inputs promise transformative outcomes across multiple domains. Enhanced conversion efficiencies and broadband phase-matching schemes will unlock octave-spanning visible combs suitable for precision spectroscopy of complex molecules and real-time atmospheric monitoring. Integration of up-conversion with electro-optic modulators and photodetectors will create monolithic optical frequency synthesizers dramatically reducing system complexity and cost.

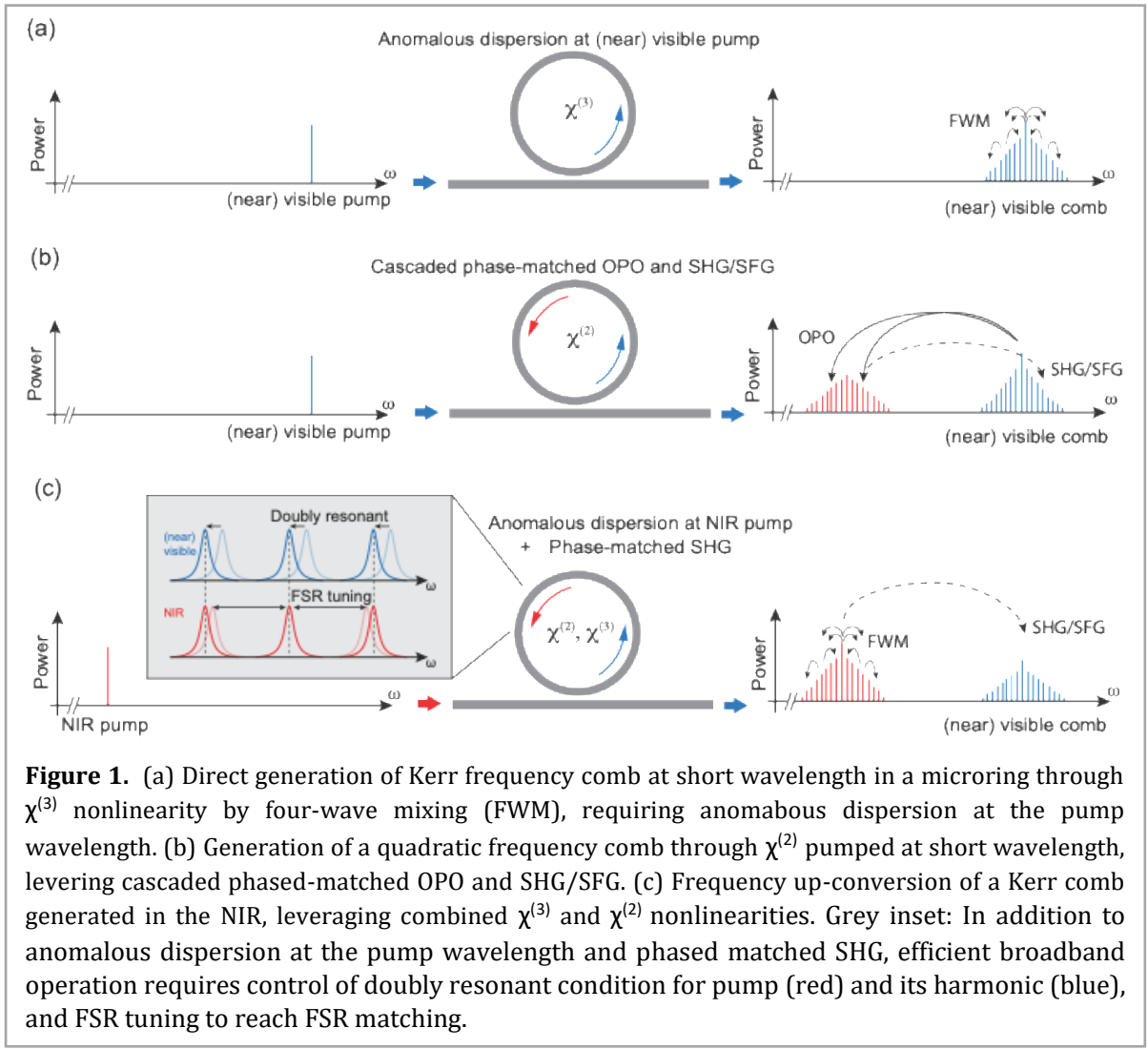

**Figure 1.** (a) Direct generation of Kerr frequency comb at short wavelength in a microring through $\chi^{(3)}$ nonlinearity by four-wave mixing (FWM), requiring anomabous dispersion at the pump wavelength. (b) Generation of a quadratic frequency comb through $\chi^{(2)}$ pumped at short wavelength, levering cascaded phased-matched OPO and SHG/SFG. (c) Frequency up-conversion of a Kerr comb generated in the NIR, leveraging combined $\chi^{(3)}$ and $\chi^{(2)}$ nonlinearities. Grey inset: In addition to anomalous dispersion at the pump wavelength and phased matched SHG, efficient broadband operation requires control of doubly resonant condition for pump (red) and its harmonic (blue), and FSR tuning to reach FSR matching.

**Current and future challenges**

Frequency conversion of telecom Kerr combs is attractive as it leverages well-established telecom technology. However, all implementations to date face fundamental challenges characterized by tradeoffs between conversion bandwidth, efficiency, and generated powers. While microring resonators are capable of highly efficient frequency doubling [9], their practical usage for broadband conversion is hindered by phase matching conditions that often constrain the conversion spectral bandwidth. Broadband SHG in a ring resonator must satisfy three requirements simultaneously: phase-matching, doubly resonant condition for pump and second harmonic (SH), and free spectral range (FSR) matching.

Perfect phase matching is usually prevented by dispersion. Workarounds include quasi-phase-matching (QPM) such as by electric-field poling of a ferroelectric waveguide [9], or intramodal phase-matching (IM-PM) [10]. As QPM by electric-field poling is highly sensitive to fabrication, most previous broadband demonstrations leveraged IM-PM, generating the SH comb on a higher-order transverse mode. This approach requires careful mode coupling and extraction schemes that can introduce additional losses and complexity. More generally, the system must efficiently couple and extract light at both the fundamental and harmonic frequencies simultaneously: optimized operation across the entire conversion bandwidth is particularly challenging in single bus designs, and most works have reported poor outcoupled power, or collected scattered light from the top of the device.

Owing to dispersion, a frequency mismatch exists between the pump and SH cavity modes. The doubly resonant condition is hence typically unachievable deterministically, even with tailored designs, as the resonance position must be controlled with a precision comparable to the linewidth — difficult to achieve due to fabrication tolerances. The most common solution is thermal tuning, leveraging the dispersion of the thermo-optic coefficient to compensate for small deviations from the doubly resonant condition. However, this method is not scalable, does not allow for precise control of the operating wavelength, and becomes impractical in systems with large FSRs.

If achieving a doubly resonant condition is difficult at a given pump wavelength, broadband SHG involving multiple resonances is even more challenging, as FSRs of pump and SH frequencies are generally different, limiting efficient conversion to only a few spectral lines. In principle, this can be mitigated by adjusting the waveguide cross-section to match the group velocity, but at the cost of dispersion engineering, which is already required for controlling the comb properties or generating soliton states.

Beyond these challenges, achieving high efficiency also places stringent requirements on the material platform. The ideal system demands materials with strong $\chi^{(2)}$ and $\chi^{(3)}$, and excellent power handling capabilities. Additionally, low optical losses are essential to maintain high quality factors at both pump and SH, maximizing the enhancement provided by the resonant cavity.

**Advances in science and technology to meet challenges**

Resonant frequency doubling in telecom Kerr comb microrings presents significant challenges that demand coordinated advances in design, control, and materials. A growing trend in integrated photonics is to move toward multi-resonator systems. These configurations offer enhanced flexibility in tailoring optical properties, crucial for advanced functionalities. Recent demonstrations have shown the power of such approach, from generating near-visible soliton microcombs [11], to realizing Vernier microcombs optimized for integrated optical clocks [12], and greater efficiency in nonlinear processes [13]. Most progress so far has been achieved for $\chi^{(3)}$ processes in silicon nitride, due to its maturity and low loss. However, exploring new materials with either enhanced $\chi^{(3)}$ or intrinsic $\chi^{(2)}$ nonlinearities, exemplified by the growing interest in thin-film lithium niobate and lithium tantalate, is key to expanding the performance envelope of integrated devices. In the context of frequency doubling, minimizing optical losses to ensure high quality factors at both the pump and harmonic frequencies, along with reliable and reproducible fabrication, remains essential.

Precise and tunable control mechanisms are especially critical to achieve doubly resonant condition and group-velocity matching. Key to this is the ability to independently tune fundamental and harmonic resonances, allowing optimization of each without compromising overall system performance. This level of control has recently been demonstrated using sets of linearly uncoupled—but nonlinearly coupled—microresonators [14]. Such architectures enable independent optimization of resonance and dispersion for each wavelength, achieving QPM over a 200 nm bandwidth and generating milliwatt-level SHG. Notably, a 12.5 THz Kerr comb was frequency-doubled with upconverted powers reaching 10 mW. Further design improvements aimed at reducing optical losses and increasing nonlinear interaction lengths are needed to approach the performance levels of single-wavelength systems.

Phase-matching strategies vary depending on the material platform. For IM-PM, future developments will likely involve sophisticated higher-order mode engineering to optimize conversion efficiency while preserving practical device architectures and achieving simultaneous phase and group velocity matching. Materials offering flexible $\chi^{(2)}$ properties—such as periodically poled structures—present promising routes for deterministic QPM while remaining compatible with scalable photonic integration platforms. Still, fabrication-tolerant structures remain a critical requirement for both broadband and narrowband frequency conversion devices, an area where lithium niobate platforms continue to face challenges [15].

Another emerging strategy is optical poling of amorphous materials via the coherent photogalvanic effect, which enables self-reconfigurable $\chi^{(2)}$ gratings [16]. While such effective $\chi^{(2)}$ strengths are lower than in traditional materials, conversion efficiencies exceeding 2500%/W [17] and milliwatt-level powers [18] have been achieved, highlighting the potential of this technique. Continued research into improving the photogalvanic coefficient and extending the nonlinear interaction length will be essential to fully unlock its capabilities.

**Concluding remarks**

Integrated resonant up-conversion of Kerr frequency combs represents a rapidly evolving field. The combination of mature Kerr comb technology with advances in nonlinear frequency conversion promises to unlock new applications and enable unprecedented levels of integration in photonic systems. As fabrication techniques continue to improve and new materials platforms emerge, we can expect continued progress toward efficient, compact, and versatile frequency conversion systems that bridge the telecom and visible domains with high fidelity and efficiency.

**Acknowledgements**

This work was funded by the European Research Council grant PISSARRO (ERC-2017-CoG 771647) and by the Swiss National Science Foundation (grant 214889).

# Soliton Dynamics in Dual-Mode and Dual-Pump Microresonators

John F. Donegan[1,*] and Hai-Zhong Weng[1,2]

[1] School of Physics, CRANN, AMBER and CONNECT Centres, Trinity College Dublin, Dublin 2, Ireland
[2] Center for Heterogeneous Integration of Functional Materials and Devices, Yongjiang Laboratory, 315202, Ningbo, China

E-mail: jdonegan@tcd.ie

**Status**

Microresonator soliton comb spectra that reach or exceed an octave span are essential for the demanding time and distance measurement applications of microcomb sources. Photonic integration of the pump laser and resonator devices shows great promise for small form factor packaged devices. The ideal soliton comb source should have (i) a low repetition rate, that can be measured electronically, < 30 GHz, (ii) a CW power of several mW, from a diode laser pump without amplifier,  (iii) a low < 30 GHz and close-to-zero offset frequency and (iv) a bandwidth exceeding an octave for self-referencing [1, 2]. Applications of microcombs include ultra-wide bandwidth communications with hundreds of channels, 1-Hz resolution optical frequency synthesizer at communication band, and optical clocks with precision exceeding $10^{-17}$ [3-5]. Octave spanning solitons have been observed in silicon nitride ($Si_3N_4$), aluminium nitride (AlN) and lithium niobate ($LiNbO_3$) platforms featuring low-loss materials that enable the fabrication of microresonators with quality factors exceeding 1 million [5-7].

Solitons arise from a balance between Kerr nonlinearity and dispersion within the microcavity [1]. However, achieving this balance in real systems is complicated by the thermo-optic effect—where the refractive index of the microresonator material changes with temperature. This makes the resonance of the cavity highly sensitive to thermal fluctuations. In practice, soliton generation typically occurs at a red-detuned regime of the pump laser—an operational point that is inherently thermally unstable [8]. This thermal instability makes it challenging to reliably initiate and sustain solitons. In many cases, the thermal effect either prevents soliton formation entirely or leads to brief soliton steps that are too narrow to support stable, long-term operation. Reliable, deterministic soliton generation is a crucial step toward realizing practical, integrated frequency comb systems for emerging applications in science and industry. To move toward practical, scalable use of soliton microcombs, it is essential to develop strategies to mitigate or compensate for thermal effects. These may include active thermal control, optimized material and resonator designs, or feedback mechanisms that dynamically stabilize the resonance condition. By ensuring stable soliton formation and long-term operation, researchers can unlock the full potential of microcombs in real-world systems.

**Current and future challenges**

Overcoming the thermal effects in microresonators is key to achieving deterministic and stable soliton generation in microcomb systems. Several strategies have been developed over the years to address this challenge. Early approaches involved rapidly tuning the pump laser wavelength across the cavity resonance to quickly reach the soliton state before thermal effects could destabilize it. Techniques such as "power kicking" and fast pump sweeping aided by an acousto-optic modulator—a combined engineering of the pump power and frequency—were also used to trigger soliton formation by momentarily changing the thermal and nonlinear dynamics [5, 9].

Recent advances have introduced more refined methods including resonator design with two close modes with the same polarization (named the dual-mode) to ensure the stable operation of soliton combs [6, 10]. In the dual-mode scheme shown in Fig. 1(a), the mode on the blue side allows soliton formation (main resonance) while that on the red side is used for thermal compensation (auxiliary resonance). During soliton formation, the cavity power decreases, resulting in a synchronous blue shift of the auxiliary resonance. With a suitable mode separation, the pump can access the effectively blue-detuned region of the auxiliary resonance, enabling the excitation of a second mode within the resonator. This additional absorbed power compensates for the reduction in cavity power in the main resonance, mitigating the thermal shift during the transition from modulation instability (MI) to the soliton state and thereby maintaining soliton stability [11, 12], as shown in Figs. 1(b) and 1(c). The dual-mode scheme can effectively extend the soliton existence range (SER) to an ultra-wide value of 32 GHz [11], which ensures long-term soliton stabilization against pump fluctuations. Dual-mode resonances with various mode separations, whose separation can be thermally controlled, were investigated for multi-solitons or soliton crystals (SCs) formation in the resonators [12, 13], thus enriching the nonlinear dynamics and improving the conversion efficiency to 42.5% for an octave spanning 3-SC. It should be mentioned that the thermal effects can also benefit from the interaction between adjacent cross-polarized modes [14].

Additionally, the use of an auxiliary laser, tuned to another cavity mode, has also proven highly effective in thermally stabilizing solitons [15]. Such a dual-pumping scheme has been widely investigated for soliton generation, soliton status switching and enables long-term soliton stability[16], as shown in Figs. 1(d)-1(f).

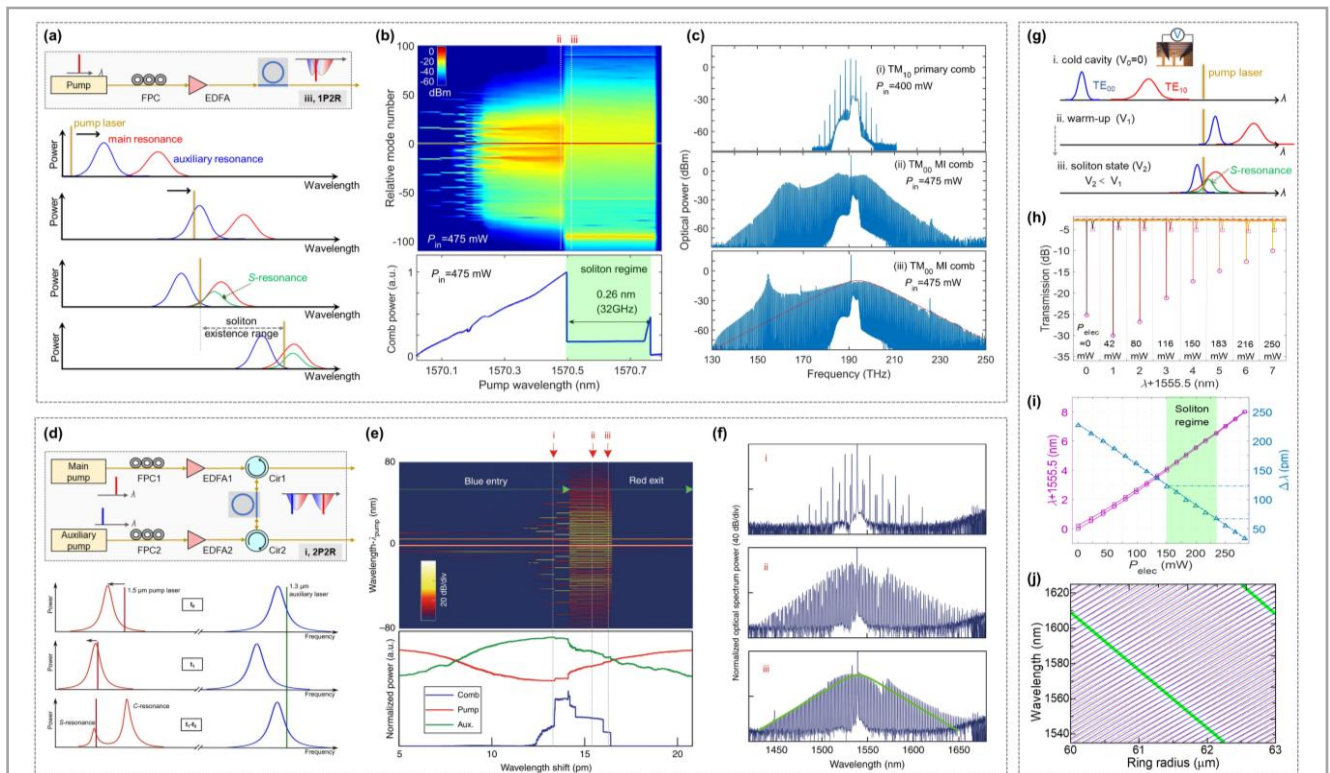

**Figure 1.** (a)-(c) Dual-mode resonators for deterministic soliton formation. (a) Diagram of the setup and principle for thermal compensation from the auxiliary mode. (b) Experimental soliton dynamics and the comb power versus pump detuning. (c) Measured microcomb spectra at specific detuning positions as shown in (b). (a)-(c) are reproduced from [11] and [12] and under a creative commons licence (CCL). (d)-(f) Dual-pumping scheme for soliton formation. (d) Diagram of the setup and principle for thermal compensation with the dual pump scheme. (e) Experimental soliton dynamics and the comb power versus pump detuning. (f) Measured microcomb spectra at specific detuning positions as shown in (e). (d) is reproduced from [15] under a CCL. (e) and (f) are reproduced from [16] under a CCL. (g)-(j) Thermally controlled resonators with improved dual-mode yield, which are reproduced by [20] under a CCL. (g) Diagram of the principle for thermally controlled dual-mode resonators for soliton generation. (h) Dual-mode resonances with near 1-FSR thermal tuning using a heater integrated 1-THz $Si_3N_4$ resonator. (i) The mode resonances wavelengths and dual-mode separation at various injected heater powers. (j). Ring radius scanning for high yield mode dual-mode crossings in C+L band.

Moreover, it was also utilized for broadening the comb spectrum and synthesizing soliton crystals [17, 18]. Much of the analysis of the dual-mode scheme for thermal compensation also applies to the dual pump scheme.

**Advances in science and technology to meet challenges**

The fabrication of microresonators as measured by the quality factor of the resonances continues to improve reaching values of $10^7$ in the best cases [19]. In turn, this will allow pump powers for achieving octave spanning combs to fall below 100 mW permitting a single diode laser pump at 1.3 μm and 1.5 μm and not requiring a bulky amplifier [3]. This looks promising for the dual pump scheme since 100 mW diode lasers that are tunable are available already and device yield issues are not a great concern here compared with the dual-mode scheme. While the dual-mode scheme has been successful in generating octave spanning solitons in AlN and $Si_3N_4$ resonators, a major challenge remained—the low yield of dual-mode devices, which heavily relies on stable and high-precision fabrication processes.

To address this, on-chip integrated heaters capable of tuning the cavity resonance by more than one free spectral range (FSR) were used [20], as shown in Figs. 1(g) and 1(h). These heaters not only relax the stringent fabrication requirements but also allow devices with sub-optimal mode spacing to be thermally adjusted into a regime where effective compensation enables soliton generation, as shown in Fig. 1(i), effectively improving device yield. Moreover, the thermal tuning method allows the transition from multi-soliton to single-soliton states, achieving simultaneous optimization of both soliton conversion efficiency and spectral bandwidth. Thermal tuning has also proven effective in precisely adjusting the comb repetition rate and carrier-envelope offset frequency, which are critical for fully stabilized combs in metrology and timekeeping applications. This thermal control strategy simplifies the entire soliton generation system by eliminating the need for pump laser wavelength tuning. Instead, deterministic soliton generation is achieved simply by adjusting the voltage applied to the microresonator, offering a much more user-friendly and integrable approach. These heaters can also be used to good effect for the dual pump scheme to shift the resonances and to access the correct pump wavelengths.

Through theoretical simulations and experimental analysis, we now propose a more efficient dual-mode design strategy: scanning the microcavity radius instead of the ring width. This approach mitigates the low dual-mode yield caused by fabrication imperfections. As illustrated in Fig. 1(j), when the pump laser operates in the C+L band, a 3-μm scanning of the ring radius enables 100% mode crossings in 800-nm thick $Si_3N_4$ microresonators with a radius of approximately 60 μm. Combined with thermal tuning, this strategy allows for the realization of a large number of dual-mode devices capable of generating soliton frequency combs. Moreover, scanning the cavity radius enables tuning of the carrier-envelope offset frequency $f_{ceo}$, making it possible to shift $f_{ceo}$ into the detection range of standard photodetectors (i.e., <30 GHz) [20].

**Concluding remarks**

Great strides have been made in generating octave spanning solitons in microresonators. There is now a clear path towards the commercial development of this technology. The list of requirements mentioned at the beginning are mostly in place. Should commercially developed resonators with quality factors of $10^7$ become available with a significant yield, this area of active research will make a large impact in photonics.

**Acknowledgements**

The authors acknowledge funding from Research Ireland under its grants, 23/EPSRC/3920 and CONNECT centre funding (13/RC/2077_P2).

# Towards mid-IR Frequency Comb Sources in Passive Photonic Platforms

Delphine Marris-Morini[1] and Adel Bousseksou[1]

[1] C2N, Université Paris Saclay, CNRS, 10 boulevard Thomas Gobert 91120 Palaiseau

E-mail: delphine.morini@universite-paris-saclay.fr

**Status**

The development of photonics platforms in the mid-infrared (mid-IR) has seen an increasing interest in the recent years. Indeed, mid-IR spectroscopy is a nearly universal way to identify chemical and biological substances and to perform non-intrusive diagnostics, especially in the so-called "fingerprint" region (wavelength from 6 to 15 µm) in which most molecules have vibrational and rotational resonances. Developing compact mid-IR spectroscopic systems, beyond 6 µm wavelength and accessible in remote areas is thus of a high interest for a variety of applications including environmental protection, healthcare, defense, or industrial monitoring. Different schemes can be proposed to obtain compact spectroscopic systems, based on (i) a wide spectral band source and on-chip spectrometer, in a similar way to classical Fourier Transform Infrared (FTIR) spectroscopy ; (ii) a tunable laser and a wide spectral range detector, in which the spectral domain is mapped by tuning the laser wavelength ; (iii) optical frequency combs and dual comb spectroscopy which is an alternative approach, allowing fast and accurate measurement of an optical spectrum with high resolution and bandwidth [1]. As shown in the other sections of the article, on chip frequency comb sources typically rely either on mode locked lasers, non-linear optical effects in resonators or electro-optical modulation. Cascade lasers are compact and flexible lasers sources operating in the mid-IR, and frequency comb has been demonstrated both with Interband Cascade Lasers (ICL) [2] and Quantum Cascade Lasers (QCL) [3,4]. In parallel, passive mid-IR photonics platforms have been developed, in which low propagation losses can be obtained, together with large third-order non-linear optics effect (Kerr effect). On chip supercontinuum generation (SCG) has been demonstrated in the mid-IR within silicon germanium [5, 6], III-V [7], while low loss chalcogenide photonic circuits also show promising results [8]. These platforms thus present a tremendous opportunity for the achievement of microresonator-based Kerr comb generation. Indeed, the combination of large non linearities with compact and high Q-factor integrated resonator can lead to ultra-broadband frequency combs, only limited by the anomalous dispersion regime of the waveguide (typically larger than 500 cm-1) which would constitute a major breakthrough in the field of mid-IR frequency combs.

**Current and future challenges**

While microresonator-based Kerr comb generation is largely developed at near infrared wavelength range, mainly based on ultra low loss silicon nitride technology, the realization of integrated Kerr comb in the mid-IR range requires to solve both scientific and technical challenges:
- To evaluate the scaling of Kerr comb generation threshold with the wavelength, it is possible to consider modulation instability threshold [9], whose dependance with the wavelength can be written as $P_{th} \sim L A_{eff} \lambda \alpha^2 / n_2$ where L is the resonator circumference, $A_{eff}$ is the effective area of the optical mode, λ is the wavelength, α are the propagation losses and $n_2$ the non-linear refractive index.

- Interestingly, the threshold power is inversely proportional to $n_2$, that is much higher in mid IR passive platforms than in silicon nitride waveguides, by typically two orders of magnitude or more [10, 11]. However the threshold power shows also (i) a linear increase with the wavelength, (ii) a linear increase with the resonator circumference and modal effective area, that both increase with the wavelength, (iii) a quadratic dependance to the propagation losses. Considering the available power from cascade sources of a few hundreds of milliwatts, and typical mid-IR waveguide properties it is possible to deduce that mid-IR Kerr comb generation requires to obtain propagation losses in integrated resonator in the order of or below 1 dB/cm, with resonator radius in the order of 100 µm (circumference around 600 µm).
- The second main challenge is related with the availability of commercially available devices for mid-IR experimental set-up, which is much more limited than at near infrared, where telecom application has driven the development of high performance experimental systems. As an example, a high speed photodetector is usually required to measure frequency comb beatnotes, but the operation speed of commercially available mid-IR photodetector is typically limited below 1 GHz where several tens of GHz are needed for few millimeters resonator lengths. Another example is related with the lack of high resolution compact optical spectral analyzer in this wavelength range, where complex and expensive FTIR or monochromator spectroscopy systems are then required to analyze optical spectra.

**Advances in science and technology to meet challenges**

Much of the ongoing research towards mid-IR frequency comb sources in passive photonic platforms focus on the development of both high performance on-chip resonators and innovative characterization techniques. As an example, to face the limitation in terms of laser tuning properties, a thermal tuning technique has been proposed and employed in Ref 12 to characterize properly the resonance linewidth. As an other example, in Ref 13, the resonator characterization is based on the measurement of the light scattered off the cavity in the vertical direction using a MIR camera on top of the sample. Regarding the resonators properties, a selection from the state of the art is reported in Table 1. Resonator performances are usually compared through the Quality factor (Q-factor), which depends both on the resonator design and on material properties. In details the measured Q-factor (loaded-Q factor) can be divided into two contributions : the intrinsic Q-factor of the stand alone resonator, and the coupling Q-factor related to the waveguide/resonator coupling. While the latest contribution depends on the coupling condition, the intrinsic Q-factor is inversely proportional to the loss coefficient and to the wavelength. It is thus not surprising that the Q-factor decrease when the wavelength increases. Interestingly it can be seen in Table 1 that high Q factor have been obtained with both III-V and Ge/SiGe based resonators, with different designs. The highest reported values are $10^6$ in the short wavelength range [13] and $2.2 \times 10^5$ in the long wavelength range in SiGe/Ge platform [14]. Recent work on InGaAs/InP showed a loaded Q factor of $6.5 \times 10^5$ with a corresponding intrinsic loss of 0.2dB/cm$^{-1}$ in the short wavelengths (1dB/cm$^{-1}$ at 8.5µm wavelength) [15]. These values can be favourably compared to the targeted values reported in the previous section. Indeed propagation losses of 1 dB/cm corresponds to intrinsic Q-factor of around $2 \times 10^5$ at 5 µm wavelength, and $1.3 \times 10^5$ at 8 µm wavelength. However, the main obstacle now for the realization of microresonator based Kerr comb generation is that in any reported demonstration, the resonator circumference is larger than 1 mm. The development of more compact structures will be essential for extending frequency comb technologies to mid-IR wavelengths.

Table 1. Selection from the state of the art : optical resonators in passive photonic platforms operating in the mid-IR range (with potential application beyond 6 μm wavelength).

|  | Ref | Wavelength | Bend radius | Loaded Q-factor |
|---|---|---|---|---|
| SiGe/Si microring | [13] | 3.5–4.6 μm | 250 μm | $10^6$ |
| InGaAs/InP racetrack | [15] | 4.6 μm | 350 μm [d] | $6.5 \times 10^5$ |
| InGaAs/InP racetrack resonator | [16] | 5.2 μm | 102 μm [a] | $4.9 \times 10^4$ |
| Ge microdisk on glass pillar coupled to partially-suspended Ge waveguides | [14] | 8 μm | 225 μm | $2.2 \times 10^5$ |
| Graded SiGe /Si racetrack resonator | [12] | 7.45-9μm | 350 μm [b]<br>250 μm [c] | $10^5$ -$1.7 \times 10^4$<br>$4 \times 10^4$ -$4 \times 10^3$ |

Racetrack resonators also include linear sections for light coupling which are (a) 150 μm long ; (b) 10 μm long ; (c) 45 μm long ; (d) 105 μm long

**Concluding remarks**

The development of mid-IR frequency comb sources in passive photonic platforms is appealing for the realization of ultra-broadband frequency combs for multipurpose sensing and spectroscopy. While significant advances have recently been made in the development of high performance on chip micro resonance, an analysis of the scaling of Kerr comb generation threshold with the wavelength indicates that currently the main challenge is to improve the compactness of the resonators, without deteriorating their Q-factor (or in other words to increase the resonator finesse). Interestingly, considering the recent progress in this domain, such improvements seem doable in the coming years. As a next challenge, innovative solutions will then have to be found regarding the control and stabilization of mid-IR sources to be used as optical pump for the driving of on chip mid-IR frequency comb sources.

**Acknowledgements**

The authors would like to acknowledge funding from the European Union's Horizon Europe research and innovation program (101128598 - UNISON) and from the European Research Council (101097569 - Electrophot). This work was partly supported by the French RENATECH network cleanroom facilities.

# Frequency Combs Produced by Femtosecond Optical Parametric Generators and Oscillators

**Markku Vainio[1]**

[1] Department of Chemistry, University of Helsinki, Helsinki, Finland

E-mail: markku.vainio@helsinki.fi

**Status**

Although several methods exist for generating optical frequency combs, direct comb generation in the mid-infrared (MIR) spectral region remains challenging. This region, especially the wavelengths from 2.5 to 12 µm, contains strong and distinct spectral features of many important molecules, making it important for gas spectroscopy and sensing. Ideally, frequency combs can provide the broad spectral coverage, high coherence and low intensity noise needed for selective multispecies gas analysis with high detection sensitivity.

A common approach to transfer optical frequency combs produced by near-infrared mode-locked lasers to other spectral regions is to use (quadratic) nonlinear optics, such as second-harmonic generation (for UV–VIS) or optical down-conversion (for MIR). Down-conversion can be achieved through difference-frequency generation (DFG), including intrapulse DFG or optical rectification [1, 2], as well as through optical parametric generation or oscillation, pumped by a single mode-locked femtosecond (fs) laser. For example, if light from a 1-µm fs laser passes through a second-order nonlinear crystal with appropriate phase-matching, signal and idler combs centered at 1.5 µm and 3 µm, respectively, can be generated (Figure 1). This process is generally referred to as optical parametric generation (OPG). When the crystal is placed inside an optical resonator, an optical parametric oscillator (OPO) is formed. The OPO can be configured as singly resonant (where either the signal or idler wave resonates) [3] or doubly resonant (where both signal and idler resonate) [4, 5]. Since the OPO is pumped by a femtosecond laser, the OPO cavity round-trip time must be synchronized with the pump laser's pulse interval.

It is worth noting that these solutions offer benefits beyond mere optical frequency conversion. For instance, degenerate doubly resonant OPOs can produce ultrabroadband comb spectra [5], and CW-seeded femtosecond OPG enables flexible comb repetition-rate adjustment, noise reduction, and simplified synchronization of two combs for dual-comb spectroscopy [6, 7].

**Current and future challenges**

The doubly resonant OPO configuration is attractive for several reasons. First, when operated at degeneracy – such that the signal and idler are indistinguishable (e.g., a 1.5 µm pump producing a half-harmonic spectrum centered at 3 µm) – the MIR output comb is automatically phase-locked to the pump comb, facilitating full comb stabilization [8]. Second, operation at degeneracy enables very broad spectral coverage, extending over an octave [5]. The main challenge is that a long and bulky OPO cavity is required to achieve the ≤1 GHz repetition rate (comb tooth spacing) often needed for molecular spectroscopy.

The singly resonant OPO offers high average output power (several watts), albeit at the cost of a high threshold power (also on the order of watts). While it can also be phase-locked to the pump comb for the generation of a fully stabilized MIR comb, the process is technically more involved [3]. Similar performance in terms of threshold, output power, and spectral width can be achieved in a simpler manner using direct femtosecond-pumped OPG without a cavity. Although the OPG process does not inherently produce an offset-stable output comb, this can be achieved by introducing an additional CW seed laser, which acts as a pre-generated comb tooth in the signal or idler region [6, 7]. However, the generated spectral bandwidths are typically limited to a few hundred nanometers. Additionally, when using bulk crystals, the required average pump power usually exceeds 1 W, similar to that of singly resonant OPOs.

In general, each of these solutions (doubly or singly resonant OPOs and CW-seeded OPG) has its own advantages and limitations. To date, no single approach offers uncompromised low-noise performance in a compact, robust implementation with low power consumption. Moreover, most demonstrations have relied on MgO-doped lithium niobate as the nonlinear material, which limits the spectral output to wavelengths shorter than approximately 5 μm. While this region is important, there is also a need to extend access to longer infrared wavelengths, reaching 10 μm and beyond.

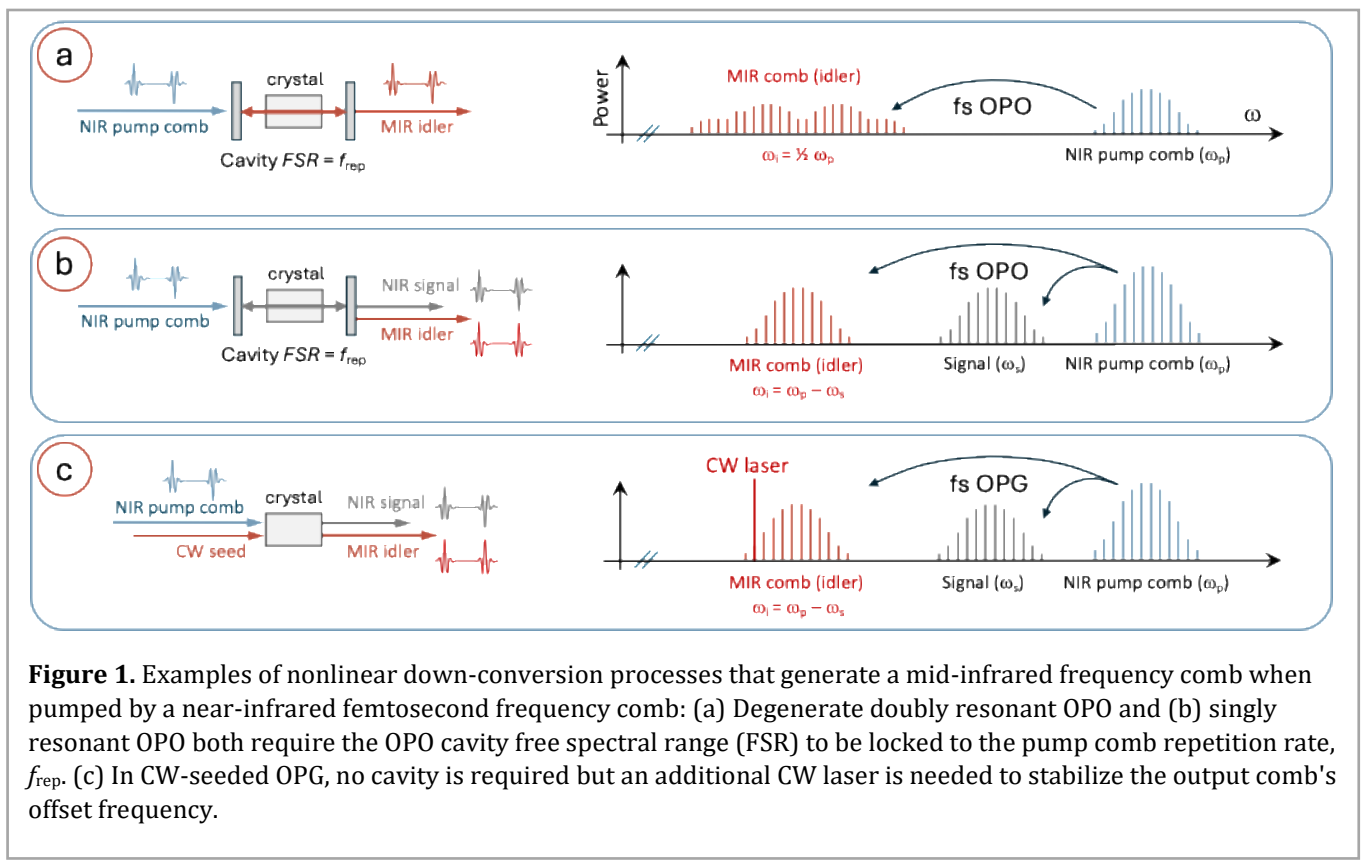

**Figure 1.** Examples of nonlinear down-conversion processes that generate a mid-infrared frequency comb when pumped by a near-infrared femtosecond frequency comb: (a) Degenerate doubly resonant OPO and (b) singly resonant OPO both require the OPO cavity free spectral range (FSR) to be locked to the pump comb repetition rate, $f_{rep}$. (c) In CW-seeded OPG, no cavity is required but an additional CW laser is needed to stabilize the output comb's offset frequency.

**Advances in science and technology to meet challenges**

Much of the ongoing research focuses on the development of solutions based on nonlinear waveguides. Although fully integrated systems (including the pump laser) are still a long-term goal, recent work on lithium niobate waveguide platforms has shown promising results towards miniaturization. For example, on-chip periodically poled lithium niobate (PPLN) resonators have been used to generate ~10 GHz repetition-rate OPO combs with broad wavelength tunability and low oscillation thresholds, requiring only milliwatt-level average power (~100 fJ per pump pulse) [9]. While instantaneous broadband output is yet to be demonstrated, the potential for dispersion engineering offers a promising path forward. Extending the length of low-loss waveguide resonators is needed to further reduce the repetition rate of OPO combs.

The significantly enhanced nonlinear interaction and the ability to control dispersion in nonlinear waveguides also provide a path to improving the performance of CW-seeded OPG comb generators. Efficient operation with low pump thresholds (mW-level average power or ~25 pJ per pulse) has already been demonstrated [10]. However, further advancements in waveguide engineering are needed to achieve ultrabroad spectral bandwidths comparable to those of degenerate OPOs using free-space resonators. Being resonator-free, CW-seeded OPG is inherently repetition-rate agnostic and could potentially be integrated with near-infrared microresonator-based pump combs. Combined with seed laser gating for repetition-rate down-conversion [6], this approach could be interesting for gas sensing applications.

Given the inherent complexity of nonlinear devices, theory development and simulations play a crucial role in performance optimization. Examples include efforts to understand temporal simultons [11] and pulse trapping [12] – both of which are important for optimizing the efficiency and spectral characteristics of femtosecond nonlinear converters.

Finally, new nonlinear materials and fabrication technologies will be essential for extending frequency comb technologies to longer wavelengths. While PPLN is excellent for targeting the 3–5 µm spectral window, many applications would benefit from access to the 7–12 µm range. This has already been achieved with bulk crystals [2, 4, 13], and progress toward low-loss waveguide devices in this range has been made with materials such as GaAs [14].

**Concluding remarks**

The hand-in-hand development of theory and experimental platforms based on dispersion-engineered nonlinear waveguides is also driving progress towards field-deployable MIR frequency comb sources operating in the 2.5 to 12 µm spectral range; an important region for spectroscopy and sensing. While compromises may be necessary, it is essential to keep in mind the stringent requirements of many use cases. For example, dual-comb spectroscopy requires low intensity noise and extremely high mutual coherence between the two combs [15].

Significant advances have recently been made with lab-scale systems, enabling accurate spectroscopic measurements that, through improved spectroscopic databases, benefit a wide range of optical sensing applications, including atmospheric and environmental monitoring [13]. Continued development of OPO and OPG technologies towards compact, low-power devices holds promise for the widespread implementation of dual-comb spectroscopy in field applications. Potential use cases include direct pollution and greenhouse gas monitoring, as well as medical diagnostics [16]. Other applications benefiting from these nonlinear techniques include random bit generation and optical computing [17].

**Acknowledgements**

The project (22IEM03 PriSpecTemp) has received funding from the European Partnership on Metrology, co-financed from the European Union's Horizon Europe Research and Innovation Programme and by the Participating States.

# Frequency Combs in High-Quality-Factor Fibre Fabry-Pérot Resonators

**Thomas Bunel, Matteo Conforti and Arnaud Mussot[1]**

[1] *University of Lille, CNRS, UMR 8523 PhLAM Physique des Lasers Atomes et Molecules, F 59000 Lille, France*

E-mail: arnaud.mussot@univ-lille.fr

**Status**

Over the past decade, Kerr nonlinear resonators have revolutionized optical frequency comb generation, enabling stable, coherent combs through purely passive cavities. Pumping a high-finesse resonator with a continuous-wave laser drives the formation of stable pulse trains leading to frequency comb spectra. Kerr combs inherently inherit the phase stability of the pump laser, avoiding active mode-locking or gain media, allowing robust and compact implementations [1, 2]. Fiber ring resonators and integrated microresonators have been the two dominant platforms so far. Fiber rings, with meter-scale cavities, have long served as versatile laboratories for exploring nonlinear cavity dynamics such as modulation instability [3], solitons [4], and switching waves [5]. Their low repetition rates and tunability make them ideal for fundamental studies. By contrast, chip-scale microresonators have enabled broad bandwidths, GHz–THz line spacing, and ultra-compact, low-power devices, spearheading comb applications in integrated photonics [1,2, 6 ]. Recently, a third platform—fiber Fabry–Perot (FFP) resonators—has emerged as a promising middle ground [8-18]. They can be short (centimeter range [8-10]) encapsulated in ferrule or longer in corded versions (Fig. 1) [11-17]. These compact cavities combine the high finesse and GHz-scale free spectral range (FSR) of microresonators with the ease of implementation of fiber-based systems (Fig. 1). Fabricated with fiber terminated by high-reflectivity Bragg mirrors, they achieve finesses of 400–1700 and repetition rates in the <500 MHz–10 GHz range. Their plug-and-play FC/PC connectors allow almost loss-less integration into all-fiber photonic systems. In addition this feature gives simple access to spatial multiplexing, by means of commercially available all fiber multiplexors developed for telecommunications applications. This extra degree of freedom enables the generation of several mutually coherent frequency combs sources because generated within a monolithic device. Uniquely, their FSR aligns with the Brillouin gain shift of standard fibers, enabling exploration of Brillouin–Kerr interactions, either for generating ultra stable combs [9,10 ] or exploring new comb formation mechanisms [12]. Note that their versatility also offer the opportunity to exploit the coupling between two cavities for self-starting pulse train generation [18]. A detailed review on recent results and future prospectives brough by these platforms is presented in Ref. [11].

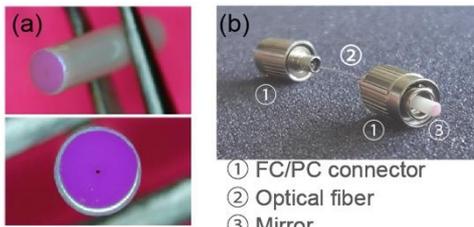

**Figure 1**. Fiber Fabry Pérot cavities. (a) Encapsulated in a ferulle [8 ] and (b) corded [11-17 ].

**Current and future challenges**

Fiber Fabry–Perot resonators have demonstrated remarkable potential for ultra stable frequency comb generation [8-11], yet several challenges and opportunities remain to fully exploit their unique capabilities. One key avenue is the control of dispersion and nonlinear interactions to increase the comb bandwidth. By selecting different fiber types within the cavity, the dispersion regime can be tuned from anomalous to normal, enabling access to a variety of nonlinear states. In the anomalous dispersion regime, we demonstrated broadband combs spanning over 24 THz with 1.2 GHz repetition rate, corresponding to ultra-short cavity solitons with estimated durations of ∼80 fs (Fig. 2 (a)) [15]. Moving to normal dispersion, switching waves can be observed using pulsed pumping. These are characterized by distinct shoulders in the spectrum, shifted up to ∼8 THz from the pump using low normal dispersion fibers (Fig. 2 (b)) [14]. An intriguing feature of these resonators is their ability to host strong interactions between the Kerr and Brillouin effects. Their centimeter-scale length produces FSRs in the GHz range, comparable to the Brillouin gain shift in silica (∼10 GHz). By precisely tuning the cavity length, it is possible to balance stimulated Brillouin scattering (SBS) and Kerr effects, enabling thermo-optic compensation effects [9] that yield remarkable comb stability. Conversely, we also demonstrated that this interplay can give rise to parametric instabilities (Fig. 2(c)), transforming a continuous-wave pump into a modulated signal [12]. Consequently, in the normal dispersion regime, the continuous wave signal may be transformed into a stable pulse train through the switching wave mechanism [12]. Such SBS-assisted dynamics open a new route to comb generation from CW pumping, complementing other mechanisms such as mode crossings and embedded cavity effects [19]. Beyond single-comb operation, fiber Fabry–Perot resonators uniquely enable multi-comb generation by exploiting spatial multiplexing. Their butt-coupled geometry and compatibility with telecom-grade all fiber components allow one to harness polarization, transverse modes, and even multicore fibers to generate mutually coherent frequency combs within a single monolithic device—without active locking. We demonstrated dual-comb operation using orthogonal polarization states [13], as well as two-core fiber cavities producing two combs with slightly different repetition rates enabling down-conversion of optical spectra into the radio-frequency domain [20].

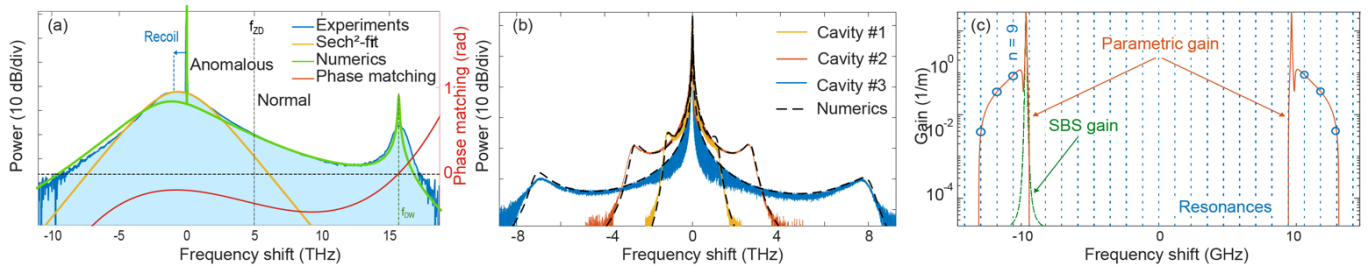

**Figure 2**. Broadband spectra and Kerr-Brillouin based phase matching processes in FFP cavities. (a) Spectrum of a cavity soliton in a highly nonlinear anomalous dispersion FFP cavity from [15]. (b) Spectra of switching waves by varying the dispersion in several FFP cavities from [14]. (c) Phase matching curve due to the SBS-Kerr balance in FFP cavities from [12].

**Advances in science and technology to meet challenges**

Meeting the challenges of fiber Fabry–Perot resonators will require advances in cavity design, stabilization, and control of complex dynamics. Future work could focus on engineering cavity dispersion and nonlinearity to extend operation beyond the telecom band, with mirror designs incorporating chirped or broadband coatings for octave-spanning combs and f–2f applications. New mirrors and exotic fibers—such as photonic crystal, tapered fibers or highly nonlinear fibers (silicium, chalcogenide for instance)-could enable operation in the visible or near-infrared, while maintaining high finesse. Improving environmental stability is another priority, through temperature-insensitive fibers (e.g., ULE glass), vibration isolation enclosures, and active stabilization using piezoelectric control. Such improvements would make Fabry–Perot combs competitive for stable microwave generation and high-resolution spectroscopy applications. On the fundamental side, the dynamics of cavities exhibiting several transverse dimensions (multimode and multicore) remain largely unexplored, offering opportunities to study the linear and nonlinear interaction to shape the comb structures, as demonstrated in microcomb platforms for instance [21]. Finally, the intriguing emergence of soliton crystals in Fabry–Perot cavities warrants deeper investigation, as these multi-solitonic regimes give access to the 10-100's GHz repetition rate [6], required for number of applications, despite the FSR is limited for a few GHz.

**Concluding remarks**

Fiber Fabry–Perot resonators have recently emerged as a powerful and versatile platform for frequency comb generation, combining high finesse, accessible GHz repetition rates, and seamless integration into fibered photonic systems. They uniquely bridge the gap between fiber ring resonators and chip-scale microresonators, offering the stability and compactness required for practical applications while retaining the flexibility and tunability valued in fundamental research. Recent demonstrations have highlighted their ability to generate broadband combs, harness Kerr–Brillouin interactions, and exploit spatial multiplexing to produce multiple mutually coherent combs. These advances underscore their potential for applications in telecommunications, spectroscopy, microwave photonics, and quantum technologies.

**Acknowledgements**

The present research was supported by the agency Nationale de la Recherche (VISOPEC, FARCO, COMBY); Ministry of Higher Education and Research; European Regional Development Fund (Photonics for Society P4S), the CNRS (IRP LAFONI); (IRP LAFONI); Hauts de France Council (GPEG project).

# Engineering Kerr Frequency Combs via Modal and Polarization Degrees of Freedom

Erwan Lucas[1], Julien Fatome[1]

[1] Laboratoire Interdisciplinaire Carnot de Bourgogne ICB UMR 6303, Université Bourgogne Europe, CNRS, F-21000 Dijon, France

E-mail: erwan.lucas@ube.fr

**Status**

Kerr frequency combs in optical microresonators—known as microcombs—have revolutionized frequency metrology and precision photonics, enabling compact and coherent comb sources for applications ranging from optical clocks and microwave synthesis to telecommunications and LiDAR [1]. These systems harness the interplay between Kerr nonlinearity and resonant enhancement to convert a continuous-wave pump into an array of equidistant spectral lines. The dominant approach relies on single-mode cavities with anomalous dispersion to stabilize dissipative Kerr solitons (DKSs), while deliberately suppressing polarization and modal couplings which are viewed as disruptive, due to uncontrolled avoided mode crossings in whispering gallery mode cavities [2,3]. This strategy, while successful, limits the diversity and controllability of generated comb states.

The theoretical foundation of Kerr combs, the Lugiato-Lefever equation (LLE) [4], describes the formation of spatial patterns in 2D cavities. It was adapted to the temporal domain in the early 1990s [5], based on the analog roles of chromatic dispersion and spatial diffraction. Subsequent predictions of polarization modulation instabilities (PMIs) in birefringent, normally dispersive fibres [6,7] hinted at a richer landscape. These vectorial effects remained largely unexplored in resonators until recently. New experiments have revealed spontaneous symmetry breaking (SSB), polarization domain walls, and polarization-symmetry-broken solitons in fibre cavities [8,9]. The recent demonstration of a chiral vectorial pulse [10] highlights the potential of nonlinear polarization coupling to structure light in regimes previously considered unfavourable.

Meanwhile, multimode nonlinear optics in fibres has introduced new spatiotemporal dynamics by exploiting transverse spatial degrees of freedom. Step- and graded-index (GRIN) fibres have enabled phenomena such as beam self-cleaning, multimode solitons, and geometric parametric instabilities in single-pass configurations [11-13]. Although theoretical studies have started to extend these dynamics to cavity systems [14,15], the field is still in its early phase.

Together, these developments signal a paradigm shift: rather than suppressing complexity, the emerging vision is to embrace and engineer it. Modal and polarization degrees of freedom, once regarded as noise sources, are now being repurposed as tools for crafting frequency combs with new functionalities.

**Current and future challenges**

Embracing polarization and modal degrees of freedom for Kerr comb generation opens a vast and largely uncharted territory in nonlinear photonics. Multimode and vectorial systems introduce intricate couplings between spatial, temporal, and polarization modes, leading to a high-dimensional solution space. These systems often exhibit strong sensitivity to initial conditions, detuning, and fabrication imperfections.

Uncontrolled interactions may induce noise, mode competition, or chaotic dynamics. The main challenge is controlling and stabilising hybrid states so that complexity serves as an asset, not a drawback.

Theoretical tools for such systems remain limited. The generalized LLE and coupled-mode approaches capture essential features but struggle when extended to highly multimode or vectorial systems. In many cases, modelling becomes intractable beyond ten transverse modes. In fibre-based cavities, the situation is further complicated by Brillouin scattering, both backward [16] and forward [17], which can either compete with or assist Kerr processes. This modelling gap hinders both fundamental understanding and device optimization.

Experimentally, controlling modal and polarization interactions requires advances in cavity design and fabrication. Small deviations from ideal geometry or material anisotropy can break resonance degeneracies or disrupt phase matching. Maintaining coherence and deterministic dynamics across interacting modes remains a critical challenge. Increasing the quality factor is crucial in multimode systems, as the larger mode volume leads to a higher nonlinear threshold.

Beyond control, new goals are emerging. A particularly ambitious frontier is the realization of spatiotemporal mode-locking in fibre resonators, potentially enabling the formation of 3D light bullets—self-localized pulses in space and time. Another is the multiplexing of multiple coherent combs within a single multimode cavity [18], offering dual frequency comb sources with excellent mutual coherence. Achieving these goals will require the co-development of theory, platform engineering, and new paradigms in resonator design.

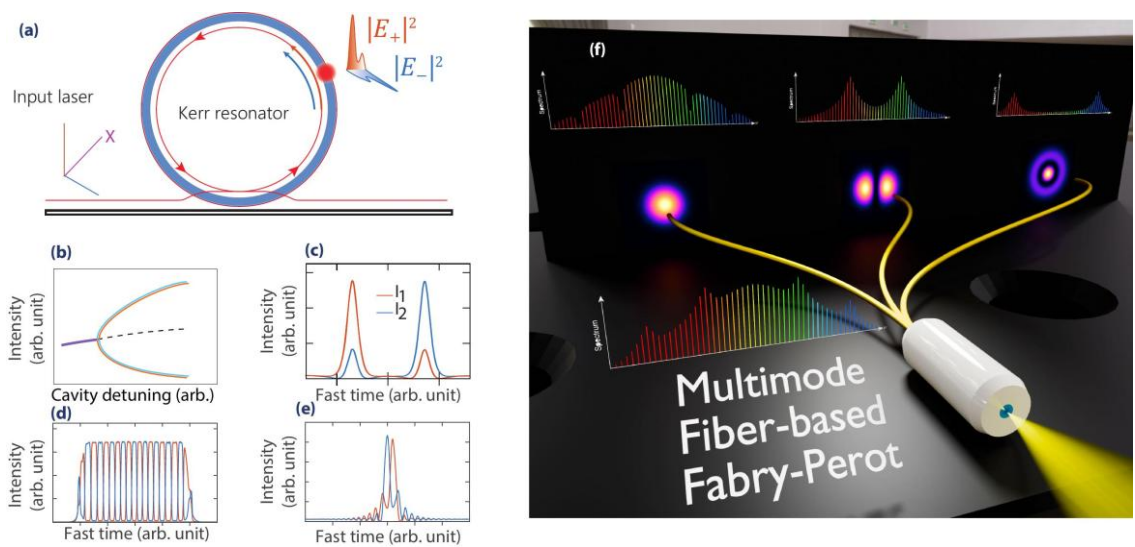

**Figure 1**: **(a)** Dual-polarization Kerr resonator; **(b)** bifurcation diagram of symmetry breaking; **(c)** vectorial dissipative soliton; **(d)** PMI pattern; **(e)** chiral vectorial pulse; **(f)** schematic of multimode dynamics in fibre resonators.

**Advances in science and technology to meet challenges**

Recent advances in modelling, fibre resonator engineering, and control strategies are laying the groundwork to tackle the challenges of complexity-enabled Kerr combs.

Improved numerical models are a cornerstone. Traditional approaches based on the generalized nonlinear Schrödinger equations (GNLSE) or modal expansions become unreliable in regimes with strong intermodal mixing. 3D+1 unidirectional pulse propagation equations—originally developed for bulk nonlinear optics—are now being adapted to fibre-based systems [19]. These models treat light as a spatiotemporal field evolving within structured media, allowing realistic simulations that account for refractive index landscapes and nonlinear coupling.

Parallel efforts are exploring data-driven approaches. Machine learning tools, particularly those for parameter space exploration and stability classification, show promise for navigating the complex dynamics of multimode systems and unveiling new regimes of operation [20].

Fibre-based cavities are emerging as ideal platforms to explore these dynamics. Their long interaction lengths, low intrinsic birefringence, and in-line configurability make them ideal testbeds. Spun fibres and integrated polarization controllers provide access to the conditions required for polarization instabilities and SSB. Graded-index fibres offer low group-velocity mismatch, enabling efficient intermodal four-wave mixing. Additionally, their larger mode volume reduces thermal noise, potentially enhancing the coherence of combs.

From an architectural perspective, Fabry–Pérot fibre cavities, where dielectric mirrors are deposited on the fibre facets [21,22], present numerous advantageous characteristics, notably facilitating access to the transverse modal structure. When combined with photonic lanterns or spatial light modulators, they offer unprecedented control over excitation conditions and modal diagnosis. These systems are ideal for generating hybridized states with tailored spectral properties and engineered coherence.

Microresonators based on photonic integrated circuits enable highly accurate manipulation of optical characteristics. Despite challenges posed by their pronounced birefringence and modal dispersion, recent advancements have introduced several mitigation strategies. New waveguide designs, polarization rotators, and mode multiplexers allow for tuneable modal dynamics [23,24]. Inverse-designed components could add further flexibility, enabling compact, low-loss, broadband manipulation of modal and polarization states. These innovations are essential for transitioning complexity-engineered combs to chip-scale platforms.

Together, these scientific and technological advances position the field to move from concept to demonstration, and ultimately to practical applications.

**Concluding remarks**

The field of microcomb research is at a turning point. What began as a search for stable and simple solitonic states is now evolving into a quest for programmable complexity. Modal and polarization interactions—once minimized—are being reimagined as design tools for shaping light in space, time, and frequency.

Fibre-based resonators, particularly those built on GRIN multimode fibres, provide a powerful testbed for these ideas. Their adaptable mode content, large mode volumes, and low-noise potential enable exploration of new nonlinear states—from polarization-symmetry-broken solitons to spatiotemporal light bullets and multiple combs. While key challenges remain in stabilizing these states and porting them to chip-scale platforms, advances in modelling, new device architectures, and data-driven discovery tools, are paving the way for the next generation of microcomb technologies. Orchestrating of complexity is key to uncover a new class of frequency comb sources, adaptable, coherent, and multiplexed.

**Acknowledgements**

JF and EL acknowledge financial support from the CNRS, International Research Project WALL-IN, the Conseil Régional de Bourgogne Franche-Comté (mobility program) and the FEDER-FSE Bourgogne 2014/2020 programs. EL acknowledges the ANR – FRANCE (French National Research Agency) for its financial support of the SMARTKOMBS project ANR-23-CE24-0005, and funding from the European Union's Horizon Europe research and innovation programme under grant agreement No 101137000.

# Laser Frequency Combs for Astronomy

Yuk Shan Cheng[1] and Derryck T. Reid[1]

[1]Institute of Photonics and Quantum Sciences, School of Engineering and Physical Sciences, Heriot-Watt University, Edinburgh EH14 4AS, UK.

E-mail: Y.Cheng@hw.ac.uk

**Status**

By precisely measuring the radial velocities of astronomical objects, high-resolution optical spectroscopy is a critical tool for contemporary astronomy: it provides the mass of distant exoplanets, delivers insights to star and galaxy formation, and even offers the prospect of revealing tiny changes in the fundamental physical constants. The astrophysical spectrographs used to acquire spectra have insufficient long term stability for the most demanding measurements, meaning that they need persistent wavelength calibration to reach the required precision. Traditional calibration relies on hollow-cathode lamps, whose atomic spectra suffer from limitations such as sparse, blended, saturated or faint spectral lines [1], or Fabry-Pérot etalons [2] that produce more regular lines but with linewidths that are too large to sample the line spread function of the highest resolution spectrographs. These limitations mean that the community is now turning to astrocombs—broadband and widely spaced laser frequency combs (LFCs) that deliver a sequence of ultranarrow, drift-free, regularly spaced optical frequencies on a selectable multi-GHz grid. Since Murphy et al. first proposed the use of LFCs for astronomical spectrograph calibration in 2007 [3], various LFC sources have been developed. Early demonstrations used solid-state lasers such as Ti:sapphire or fibre lasers [4,5]. but over the past decade more versatile platforms like electro-optic modulation (EOM) combs and micro-resonator-based combs have gained attention [6,7]. These sources allow more flexible control over mode spacing in the multi-GHz range, making it easier to match the resolution requirements of spectrographs. While each LFC technology has its own strengths and limitations, commercially deployed astrocombs have predominantly employed modelocked fibre laser LFCs, based on an architecture using Fabry-Pérot filtering to increase the mode spacing from sub-GHz to multi-GHz, followed by amplification then nonlinear spectral broadening, typically from the blue-green to the near-infrared (for example, [8]). Astrocombs have enabled few cm s$^{-1}$ radial-velocity calibration precisions [4,9], equivalent to a fractional optical frequency precision of ≤10$^{-10}$, as well as tracking spectrograph drifts [5] and calibrating the line-spread function (LSF) of the spectrograph sensor [10]. These various capabilities enable the long-term spectrograph calibration needed for extreme precision radial velocity measurements [11].

**Current and future challenges**

From a user perspective, the key requirements for astrocomb systems are: spectral bandwidth—sufficient to cover the operating range of the spectrograph; line spacing—optimally at ~3Δ$\lambda$, where the resolving power of the spectrograph is $R=\lambda/\Delta\lambda$; frequency accuracy—normally ~10$^{-11}$ fractional accuracy in one second, based on using a GNSS reference; linewidth—preferably ≪Δ$\lambda$ so that each comb line samples the spectrograph line spread function precisely; flatness—with a preferred fluctuation across each order of <4 dB; sufficient intensity—so that calibration spectra can be obtained without long averaging times; and reliability—to minimise costly down time of the calibrator during observations. Since many of these requirements are spectrograph specific, no one-size-fits-all solution exists, and a key challenge is to match the astrocomb characteristics to the spectrograph, with modern instruments like ANDES [12] covering very broad bandwidths at high resolutions ($R$≥100,000).

Addressing these user requirements encounters a number of specific technical challenges. The modelocked lasers conventionally employed for astrocombs produce pulses with insufficient spectral bandwidths and mode spacings to be directly useful. Instead, two additional operations—nonlinear spectral broadening and comb-line-filtering—are needed to reach the necessary multi-100-nm bandwidths and ≥10 GHz mode spacings. Filtering is normally implemented using an air-spaced Fabry-Pérot etalon, but reduces the peak power of the pulses to a level that requires strong amplification for high-repetition-rate nonlinear spectral broadening to be successful, in turn requiring highly nonlinear media (typically tapered photonic crystal fibres or waveguides) that are resistant to damage when pumped at multi-Watt levels. Conversely, broadband supercontinuum generation is readily achieved with the direct output of a modelocked laser such as Ti:sapphire [13], but subsequent filtering requires Fabry-Pérot mirrors with precisely balanced group-delay-dispersion characteristics [14], while maintaining enough reflectivity to achieve sufficient finesse to suppress primary LFC modes. Both approaches encounter the issue of comb-line sub-set ambiguity, where identifying the absolute mode number of the filtered comb line requires integrating a cw laser with a known frequency into the filtered LFC as a wavelength marker, or using an auxiliary measurement such as pre-calibration using ThAr lamp. Emerging astrocomb schemes using EOM combs [10] or micro-resonators [7] eliminate the need for filtering, but higher phase noise has been observed to considerably broaden comb lines at the edges of the spectrum.

Nonlinear broadening typically results in large variations in intensity across the spectrum which impacts the calibration precision. Spatial light modulators (SLM) are therefore needed to achieve dynamic spectral flattening, but reduce the astrocomb power considerably and potentially introduce other instabilities [15]. Reaching UV and blue wavelengths directly from third-order supercontinua processes is difficult, and requires either a second-order conversion step [16] or multi-harmonic generation in nanophotonic waveguides with the presence of gaps [17] or overlapping harmonics of different comb offsets.

Flexibility in the astrocomb output can be advantageous, for example the desire to tune the astrocomb offset has been noted by astronomers [18], but is difficult to provide using modelocked-laser astrocombs. Different spectral regions also require different astrocomb mode spacings, with shorter wavelengths needing wider spacings because the frequency resolution of a grating spectrograph is $c/R\lambda$. Also, the dense mode spacing needed to provide the best wavelength calibration is tensioned with a requirement for sparse astrocomb mode spacings suitable for line-spread function measurement.

Finally, all astrocombs must address the environmental challenges of operation at telescopes located at altitude in isolated locations, where automated and / or remote operation must be part of the design to ensure uninterrupted service.

**Advances in science and technology to meet challenges**

Efforts to improve the flexibility and simplicity of astrocombs are currently centred on the application of nanophotonic nonlinear devices, such as Si3N4 or Ta2O5 waveguides, which can be driven by EOM combs with 10 GHz or 20 GHz spacings [21], and work to understand and improve their phase noise is continuing [19]. The inherent tunability of EOM combs, in which the comb spacing is determined by the microwave drive frequency, allows the comb modes to be tuned across the spectrograph sensor to provide a characterisation of every pixel [10], and similar approaches with Ti:sapphire astrocombs are in development by using a combination of Fabry-Pérot and repetition-rate tuning [20]. The extension of single supercontinua to the UV range is possible by using simultaneous second- and third-order nonlinear processes in nanophotonic thin-film lithium niobate (TFLN) and has provided 330–2400 nm coverage [21], however regions of overlapping harmonics cannot be directly used for astrocombs unless the pump pulses are controlled to have a carrier-envelope offset frequency of zero, which is technically straightforward but as yet undemonstrated in this context. Improved methods for astrocomb flattening and control are also in development, in order to maintain a constant photon flux per comb mode[22] and to provide line-by-line control for greater flexibility [23].

**Concluding remarks**

Astrocomb technology has developed considerably in the last decade, both in terms of the prevalence and reliability of commercial systems and the emergence of new technology platforms based on nanophotonic frequency conversion that promise simpler and possibly more robust systems. A remaining question is to what extent astrocombs can improve spectrograph calibration beyond what is already possible by using conventional atomic and molecular references such as ThAr lamps and iodine cells. Fundamentally, the starlight impinging the spectrograph sensor follows a different optical path from the astrocomb calibration light. At the few cm s-1 level—where astrocomb calibration will be most valuable—the precision limits imposed by the subtly different wavefronts of the science and calibration light may ultimately clamp the performance at a level comparable to classical calibrators. In this limit, only a truly single-mode photonics telescope, enabled by extreme adaptive optics, would be able to fully exploit the capabilities of modern astrocomb calibrators.

**Acknowledgements**

The authors acknowledge funding for this work from the Science and Technology Facilities Research Council under grants ST/V000403/1; ST/Y001273/1; ST/X002306/1; ST/X002845/1 and ST/W005468/1.

# Microresonator-Filtered Lasers: Opportunities and Challenges

Alessia Pasquazi[1], and Marco Peccianti[1]

[1] Emergent Photonics Research Centre (EPicX), Dept. of Physics, Loughborough University, Loughborough, LE11 3TU, England, UK

E-mail: a.pasquazi@lboro.ac.uk

**Status**

Microresonator-filtered laser technology offers an architecture for ultrafast pulsed sources to overcome gain bandwidth limitation and leverage the virtually unlimited spectral conversion of nonlinear parametric interaction(1–3). This is enabled by decoupling the optical path in the gain medium from that of the parametric resonator, a distinctive feature of this platform, typically implemented via a nested cavity design (Fig. 1).

Crucially, the amplifying cavity does not require any ultrafast nonlinear saturation element to sustain pulsing. Instead the regime can exists in a purely linear amplifying cavity, dissipatively coupled with a Kerr resonator. This is a defining difference from classical passive mode-locking designs, such as additive-pulse mode locking, Kerr-lens mode locking, or figure-eight lasers and its derivatives, where Kerr elements act as ultrafast saturable absorbers and the optical spectrum is bounded by the gain bandwidth.(4,5) While the gain limits the recirculating pulse in the amplifier, the bandwidth of the pulse in the nonlinear Kerr cavity is eventually defined by the its energy and the group velocity dispersion, as in systems governed by dissipative variations of the nonlinear Schrödinger equation, with the Lugiato-Lefever model as a primary example(4,5). Figure 1x shows an oscillation 60 nm sustained by a gain bandwidth of less than 10 nm. This design enables naturally synchronised pulsed pumping of the microcavity, which is a popular method to enhance the bandwidth(6,7), and leads to high spectral efficiency, with demonstrated conversion efficiencies exceeding 50%(8).

Lasing spontanously emerges within the resonant lines of the nonlinear microresonator, and leverages intrinsic stabilising feedback effects enabled by, slow, energy-dependent nonlinearities, such as thermal effects and gain index changes(9), in both the amplifying and nonlinear cavities(3,10). This provides distincive self-starting and self-recovery features to the architecture without external-control. Similar stabilisation mechanisms are well explored also in continuous-wave configuration(11,12), like in thermal and self-injection locking, but here thet act on the full pulse. The balance of these nonlinearities can be precisely controlled by external global parameters (e.g., pump power, cavity length), allowing the system to naturally evolve into soliton, Turing, or crystal states without resonance scanning or complex "writing" procedures. The result is a highly versatile and robust system that can switch between different nonlinear states, including single soliton operation, simply by tuning static, global parameters(3). The coexistence of different nonlinear states is possible and does not disrupt soliton formation; on the contrary, it can add further robustness, mediated by nonlocal nonlinearities(13).

Finally, the multi-modal nature of microresonator-filtered laser fundamentally allows controling with some degree of independence the lasing modes and, by design, the accessible nonlinear states. Very recent results show the capability to control independently the lasing threshold of different soliton families and the operation of molecules and topological states, opening perspectives for topological protection and robust states and new regimes of photonic dynamics(14).

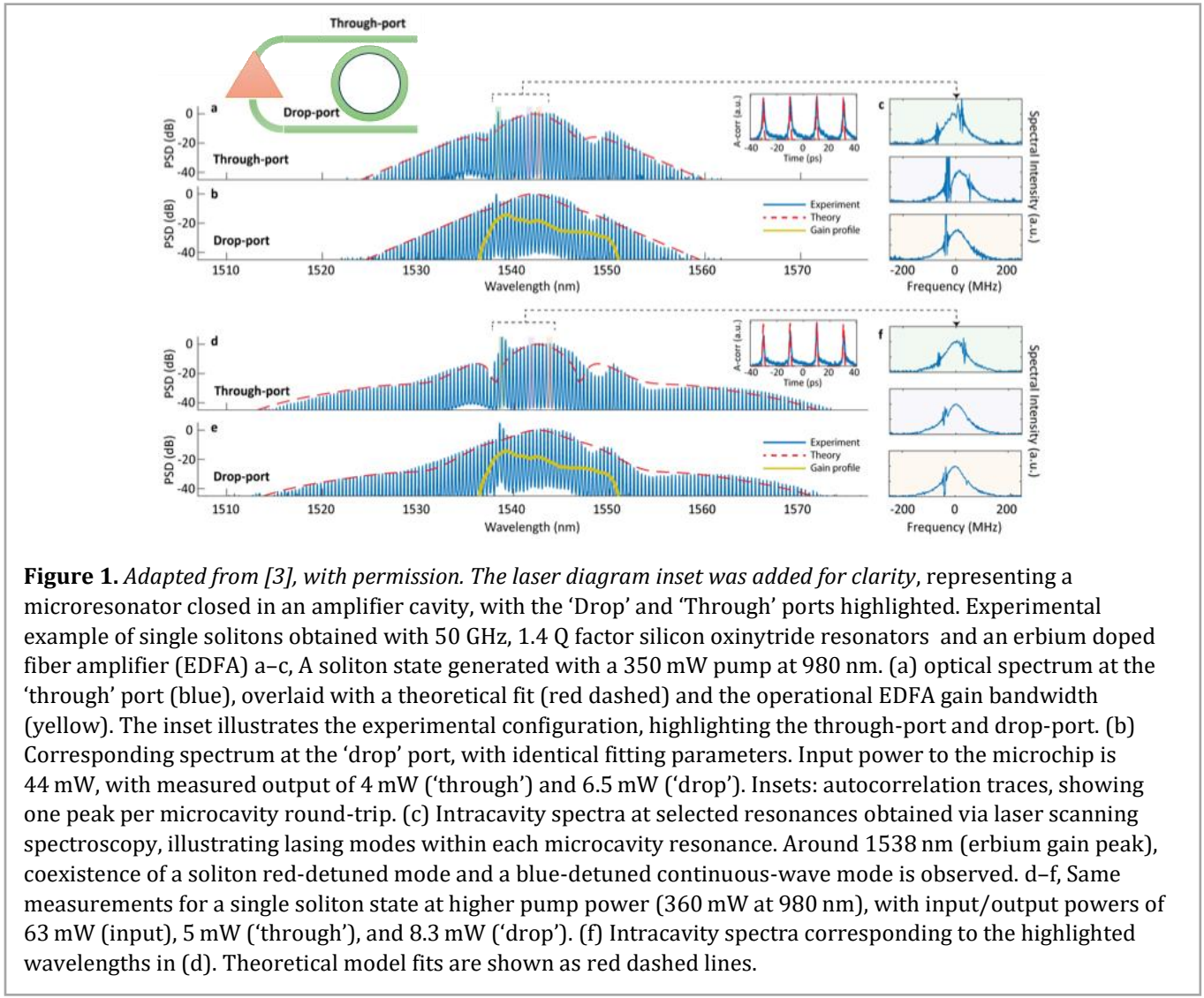

**Figure 1**. *Adapted from [3], with permission. The laser diagram inset was added for clarity*, representing a microresonator closed in an amplifier cavity, with the 'Drop' and 'Through' ports highlighted. Experimental example of single solitons obtained with 50 GHz, 1.4 Q factor silicon oxinytride resonators and an erbium doped fiber amplifier (EDFA) a–c, A soliton state generated with a 350 mW pump at 980 nm. (a) optical spectrum at the 'through' port (blue), overlaid with a theoretical fit (red dashed) and the operational EDFA gain bandwidth (yellow). The inset illustrates the experimental configuration, highlighting the through-port and drop-port. (b) Corresponding spectrum at the 'drop' port, with identical fitting parameters. Input power to the microchip is 44 mW, with measured output of 4 mW ('through') and 6.5 mW ('drop'). Insets: autocorrelation traces, showing one peak per microcavity round-trip. (c) Intracavity spectra at selected resonances obtained via laser scanning spectroscopy, illustrating lasing modes within each microcavity resonance. Around 1538 nm (erbium gain peak), coexistence of a soliton red-detuned mode and a blue-detuned continuous-wave mode is observed. d–f, Same measurements for a single soliton state at higher pump power (360 mW at 980 nm), with input/output powers of 63 mW (input), 5 mW ('through'), and 8.3 mW ('drop'). (f) Intracavity spectra corresponding to the highlighted wavelengths in (d). Theoretical model fits are shown as red dashed lines.

**Current and future challenges**

The multi-modal nature of microresonator-filtered lasers distinguishes this approach from other microcomb based on continuous-wave pumping(15). In particular, the ability to select which modes reach threshold is enabled by the microresonator acting as both a nonlinear element and a spectral filter. Remarkably, it opens access to a wide variety of nonlinear states, including those with potential topological robustness(14), which are promising for ultra-low-phase-noise metrological sources.

To leverage this property, it is necessary however to limit the presence of multiple lines within the microresonator linewidth cavity, which means minimising the ratio between the microcavity linewidth and the amplifying cavity. This places specific technical requirements on the Q-factor and the amplifier length, with a trade-off on its maximum gain especially for fibre amplifier implementations. A correct management of these parameters allows reaching high-energy or highly nonlinear states that typically require large intracavity detuning. Achieving metrological-grade emission requires full synchronisation and stabilisation of the two degrees of freedom present in this dual-cavity system. Established techniques, such as active control of the external cavity using piezo stretchers or electro-optic modulators, are readily available and allow for practical comb stabilisation. In this scenario, modern research directions focuses on approaches to improve the inherent performance via advanced approaches—including topological protection, Kerr-induced synchronisation(16), and injection locking(11)—as promising pathways that could further enhance performance. The feasibility of these methods in the platfrom requires further explorations.

Still within the stabilisation framework, slow, energy-dependent nonlinearities are an essential aspect of achieving locking and robust operation. Erbium-doped fibres, for example, naturally provide slow gain dynamics that support self-starting and stability, while the nonlinear response in semiconductor-based materials still needs to be studied in this regard and may require further engineering and optimisation. This aspect is particularly relevant for fully integrated platforms.

The platform is highly versatile with respect to resonator material systems—demonstrated in silica based glass(1), silicon nitride(17), and fibre-based resonators(18)—and erbium-doped fibres remain effective for amplification. Such hybrid fibre–integrated resonator architectures offer low energy consumption and accessible external cavity control (for example, via intracavity modulation or piezo stabilisation), delivering good output energy and spectral performance with compactness suitable for many applications. Achieving full monolithic integration remains an important perspectivefor future development.

Reliance on global parameter control for state selection places a premium on precise engineering, particularly regarding packaging and minimisation of fixed intracavity losses, as these losses directly affect the effective relative energies between the two cavities and serve as critical control parameters.

**Advances in science and technology to meet challenges**

A crucial aspect in dissipative systems is understanding the landscape of states that can be produced with a given technology. For microresonator-filtered lasers, progress in the area of multimode and multistate nonlinear interaction is key. The recent demonstration that interleaved detuning configurations can lead to a topological molecule opens an important perspective on how phenomena such as symmetry breaking affect the nature of the state and the consequences for technology—especially for the topological protection of temporal states, which has significant impact on the metrological quality of the source.

Metrology is a central aspect of this technology. Together with the implementation of topological protection, the feasibility of Kerr-induced synchronization and similar approaches to locking the system via nonlinear coupling to external sources is certainly an important capability. Its feasibility is reinforced by recent demonstrations showing that nested-cavity lasers are compatible with nonlinear mixing and synchronization using external continuous-wave signals. More broadly, progress in the current technology toward higher TRLs in terms of packaging, fabrication, and cavity engineering, will enable improved performance in the implementation of active cavity control and stabilization techniques.

At the same time, another important aspect is how slow nonlinearities affect the states that are effectively defined by the ultrafast nonlinear interactions within the cavity. This is particularly important as the source is highly multistable; phenomena such as hysteresis need to be better understood, which can not only lead to improved state selection but also enable switching and readdressing functionalities in time, with potential impact from signal shaping to more advanced computing approaches.

A more general question is what types of slow nonlinearities can support, for instance, soliton operation and formation—potentially leading to significant advances in engineering materials and integration approaches. Toward integration, research on fabrication and materials—including hybrid integration of lasing materials (III–V semiconductors, rare-earth doping) over CMOS-compatible nonlinear platforms, planar fabrication approaches for different nonlinear glasses, or technologies for quadratic nonlinearities—will certainly be beneficial. In terms of design, it will be important to address issues such as optical isolation, potentially through gain–loss-induced directionality or Fabry–Perot configurations.

Finally, microcombs are currently considered as key to create millimetre wave and THz system with unprecedented performances. The evolution of key noise features and high spectral efficiency will enable direct driving broadband emitters for WDM communications and positioning and time-domain spectroscopy systems for sensing. Regarding millimetre wave generation with laser-cavity solitons generated by microresonator filtered lasers, preliminary results(19) show compatibility with direct use in photoconductive THz antennas and time-domain spectroscopy systems.

**Concluding remarks**

Microresonator-filtered laser technology represents a distinctive approach to broadband pulsed sources. Within the microcomb field, it offers unique strengths and presents challenges that demand a tailored physical understanding. Continued progress in the fundamental physics, especially regarding slow nonlinearities, multistability, and the interaction with external fields, together with sustained innovation in materials, integration, and packaging, will be essential to deliver scalable and truly application-ready devices. Hybrid fibre–microcavity implementations have already achieved higher technology readiness levels, and will further benefit from advances in multimode fibres and spatially multistate selection in free-space cavity systems, with potential to advance both practical performance and the underlying physics, including the realization of topological operation. For progressing toward full integration, it will be essential to deepen our understanding of materials and design, as well as the role of slow nonlinearities. These advances will have significant impact in metrology, compatibility with quantum technologies, communications and sensing, and position, navigation, and timing applications.

**Acknowledgements**

This roadmap summarises the work at the Emergent Photonics Research Centre.

# Dissipative Solitons and Optical Frequency Combs in degenerate cavity VECSELs

M. Giudici[1], M. Marconi[1], A. Bartolo[1,2], N. Vigne[2], B. Chomet[2], A. Garnache[2], G. Beaudoin[3] and I. Sagnes[3].

[1] Université Côte d'Azur, CNRS UMR7010, Institut de Physique de Nice, 06200 Nice, France
[2] Institut d'Electronique et des Systèmes, CNRS UMR5214, 34000 Montpellier, France
[3] Centre de Nanosciences et de Nanotechnologies, CNRS UMR 9001, Université Paris-Saclay, 91120 Palaiseau, France

E-mail: massimo.giudici@univ-cotedazur.fr

**Status**

The customization of the spatio-temporal structure of light is a central challenge in modern photonics. Multimode photonics [1–5], a rapidly advancing field, focuses on generating and controlling complex light states for applications in information processing, photonic computing, sensing, and imaging. Lasers and driven optical resonators emitting localized structures (LS)—also known as dissipative solitons—represent promising sources for creating structured and controllable light fields [6–8]. In the spatial domain, these structures (spatial LS, or SLS) appear as individually addressable light peaks across the transverse cross-section of a resonator [9], whereas in the temporal domain (temporal LS, or TLS), they correspond to individually addressable pulses circulating within the cavity [10].

To date, SLS and TLS have been realized in two distinct classes of optical cavities. SLS require large Fresnel number cavities with a single longitudinal mode available, while TLS rely on cavities with many longitudinal modes—implying a low Fresnel number and the presence of only a single transverse mode. Degenerate cavities offer a promising route to bridging these regimes, as they allow for high Fresnel numbers in resonators with extended longitudinal dimensions, thereby enabling the generation of spatio-temporal LS and comprehensive control over the light field structure.

We have recently developed a III–V semiconductor-based vertical external-cavity surface-emitting laser (VECSEL) capable of emitting TLS in a nearly degenerate cavity configuration [11]. Operating in the near-infrared (NIR) around 1 μm, the device comprises a gain mirror and a saturable absorber mirror (SESAM). Our VECSEL platform utilizes optical pumping via commercial GaAs-based 800 nm laser diodes and beam-shaping optics, achieving homogeneous excitation over large areas (>100 μm diameter), low defect density, minimal strain anisotropy, and reduced thermal impedance. Efforts are ongoing to extend the emission range toward telecom wavelengths.

TLS generation in our system arises from multistability between various mode-locked states and the off-emission state [12], allowing for a tunable number of pulses circulating in the cavity. The gain mirror and SESAM are positioned in conjugate planes, forming a self-imaging external cavity. This configuration facilitates spatial and temporal addressing of LS by superimposing local and pulsed signals onto the flat-top stationary pump. However, the optical pump induces thermal lensing, which must be carefully managed when designing the self-imaging cavity.

The device exhibits emission of temporally localized, non-homothetic patterns. Depending on the external cavity parameters, these patterns emerge from either a Turing instability [11] or from interference between tilted beams propagating within the cavity [13,14]. In both cases, residual spherical aberrations of the optical elements play a critical role in pattern formation. Despite the nearly degenerate cavity design, the VECSEL output remains spatially correlated across the entire pumped area.

**Current and future challenges**

The results outlined above highlight the significant role of spherical aberrations in optical elements when operating near self-imaging conditions. These aberrations influence the ability to observe spatially decorrelated TLS, and even more so, the realization of spatio-temporal LS. Nevertheless, it is possible to induce multiple decorrelated emission channels—or "hot-spots"—by spatially modulating the net gain profile on the gain mirror (or on the SESAM). This strategy builds on the principle that a nearly degenerate cavity can support lasing profiles of arbitrary shape [3].

To implement this, hot-spots were engineered directly onto the top facet of the gain mirror using a sub-wavelength, non-diffractive chromium absorptive mask that locally shapes the net-gain landscape [15]. This mask was fabricated via post-growth electron-beam lithography and lift-off, enabling arbitrary patterning with 5 nm spatial resolution. As a proof of concept for spatio-temporal light control, a net-gain landscape comprising two spatially separated circular regions was realized (Fig. 1). Each region functioned as an independent TLS source, and we demonstrated spatially selective addressing of individual pulses circulating in the cavity [16]. By controlling the number of pulses per roundtrip emitted by each spot, which ranges from zero (no emission) to a maximum value $N_{max}$ determined by the ratio between the cavity roundtrip $t_{rt}$ and $t_g$, the VECSEL light emission can be customized in many different spatio-temporal states.

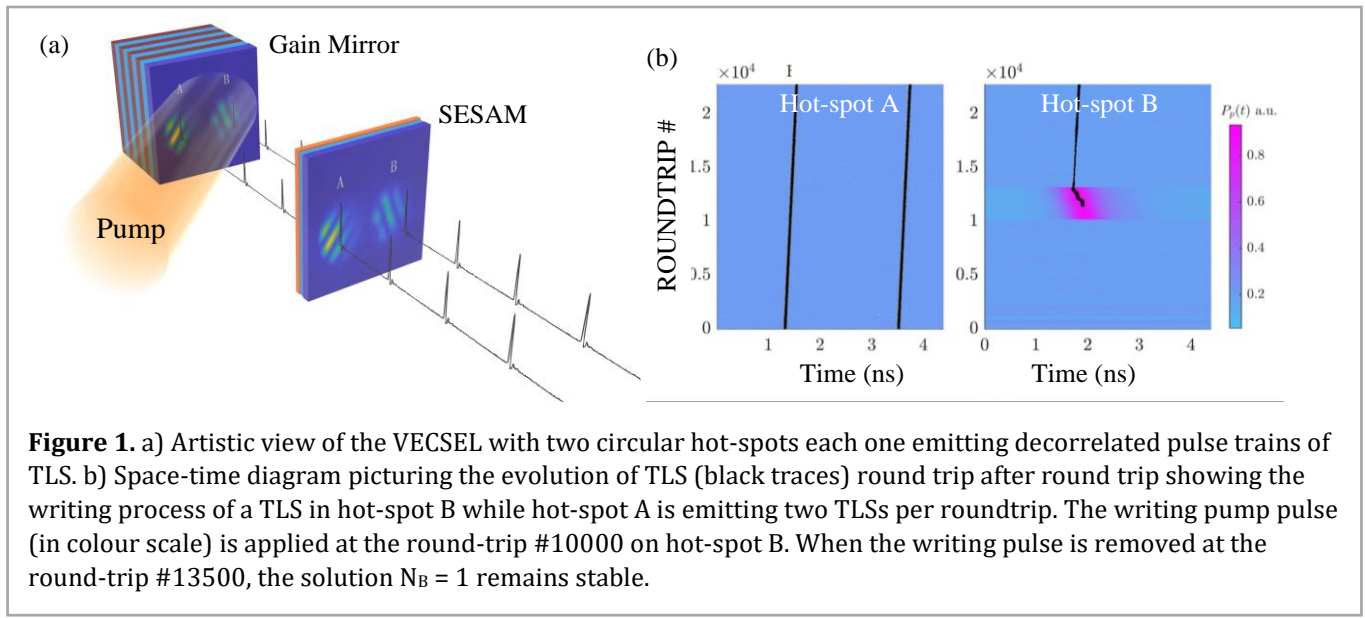

**Figure 1.** a) Artistic view of the VECSEL with two circular hot-spots each one emitting decorrelated pulse trains of TLS. b) Space-time diagram picturing the evolution of TLS (black traces) round trip after round trip showing the writing process of a TLS in hot-spot B while hot-spot A is emitting two TLSs per roundtrip. The writing pump pulse (in colour scale) is applied at the round-trip #10000 on hot-spot B. When the writing pulse is removed at the round-trip #13500, the solution $N_B = 1$ remains stable.

This development marks a significant step toward reconfigurable laser platforms capable of generating light fields with scalable complexity. By identifying a light state with the number of pulses emitted by each hot-spot, the number of different light states is given by $(N_{max}+1)^n$, being *n* the number of hot-spots. In terms of information processing, by using each pulse for carrying one information bit, the VECSEL can be considered a spatio-temporal all-optical buffer with a memory size given by the product $nN_{max}$. Another promising application lies in the generation of optical frequency combs (OFCs) from each hot-spot. In this scenario, the VECSEL enables multiplexing of OFCs that share the same active and passive media—and therefore the same noise sources—thereby enhancing mutual coherence, a critical requirement for multi-comb operations [17,18]. The repetition rate of each OFC can be tuned either by injecting a localized continuous-wave beam into the corresponding hot-spot, as shown in [16], or by modifying the geometrical parameters of the absorptive mask.

**Advances in science and technology to meet challenges**

To achieve spatio-temporal LS (also called Light-Bullets) requires engineering of lens aberrations and/or use of spatial light modulator to shape properly the wavefront profile of the intracavity electromagnetic fields. These are future challenges for mitigating the effects of optical elements aberrations. On the other hand, the performances of the degenerate VECSEL with an integrated net-gain landscape can be improved by increasing $N_{max}$ and n. While the former requires an increasing of the external cavity size or faster semiconductor gain media ($t_g$), the latter is obtained by increasing the number of hot-spots optically pumped. Several technological steps are required to achieve these goals because, at present, there is a critical low limit in the size of the hot-spots for reaching the laser threshold. While the total pump area can be increased relatively easily, there exists an upper bound on the total pump power that can be injected into the gain mirror. Overcoming this limitation will require improved thermal management, such as the integration of diamond heat spreaders into the gain section to enhance heat dissipation.

Regarding the application to multiple optical frequency combs (OFCs), further validation is needed. Specifically, measuring the mutual coherence and tunability of the repetition rate in individual hot-spots is crucial to assess their suitability for multi-comb systems. To simplify system complexity, future efforts will also explore integration of the saturable absorber directly onto the gain mirror, resulting in a monolithic MIXSEL (mode-locked integrated external-cavity surface-emitting laser) architecture [19,20].

Degenerate cavity VECSELs prove to be ideal platform for generating a fully developed spatio-temporal complexity of light states. In the future this platform will be tested for generating spatio-temporal chaos, optical turbulence and spatio-temporal excitability.

**Concluding remarks**

Degenerate-cavity III–V semiconductor VECSELs incorporating an intracavity saturable absorber are a promising platform for generating light fields with complex spatio-temporal structures. While addressable, mode-locked pulses with non-homothetic spatial profiles have already been demonstrated, the experimental realization of spatio-temporal dissipative solitons remains an open challenge. Achieving this goal will require further progress in intracavity wavefront control to mitigate aberrations introduced by optical elements.

Nonetheless, spatially decorrelated TLS sources have been successfully implemented via the introduction of a tailored net-gain landscape on the gain section. Future efforts will focus on advancing this platform for practical applications, including optical information processing, programmable spatio-temporal light generation, and the multiplexing of optical frequency combs.

**Acknowledgements**

This work was supported by the French RENATECH network. We acknowledge support from ANR (Blason and Kogit projects) and Région PACA.

# Developing Talent and Skills for Timing Technology Research and Implementation

**Richard Burguete[1] and Sarah Hammer[2]**

[1] Postgraduate institute for Measurement Science (PGI), National Physical Laboratory (NPL), Teddington, UK
[2] Science & Technology Capability, Defence Science and Technology Laboratory (Dstl), Salisbury, UK

E-mail: Richard.burguete@npl.co.uk

**Status**

Optical frequency combs (OFCs) are a key component in systems that generate high-precision timing signals. As research in the area advances and the technology develops, these systems will be adopted more readily for applications in, amongst others, aerospace, defence, healthcare, and power systems. The importance of OFCs is therefore recognised as having a high potential impact in the generation of timing signals.

The provision of precise, reliable, and resilient timing signals is critical to the UK's national infrastructure [1] as it underpins our everyday lives from communications and transport systems to computer networks and many other essential systems. Developing technology to generate assured time that is trusted, provides confidence in our technological infrastructure and ensures that it is resilient and works reliably. In addition, this will drive new investment and export opportunities that will enhance the UK's prosperity and quality of life.

Consequently, the need to develop, broadly speaking, the capability in academia for research and training and in industry to implement and exploit these technologies, is reliant on establishing a skilled and suitably qualified workforce, and one that is sufficiently workforce to execute this ambitious UK strategic programme. The main initiatives that have driven the development of a high-level workforce (doctorate level and above) are the EPSRC* Quantum Centres for Doctoral Training [2] and the UKRI Quantum Hubs [3] announced in 2023 and 2024 respectively. In addition to this, an overarching strategy that focuses on the specific needs of the wider quantum technology sector (of which timing is an important sub-set) is comprehensively set out in the UK Quantum Skills Taskforce report [4] published in May 2025.

Set within this broader context, a community of interest in the area of timing, in addition to positioning and navigation (PNT) is being established to address many of the requirements for the development of research and skills in this area, for example the Royal Institute of Navigation provides specialist training courses in this area [5]. The evolution and growth of this community, alongside the establishment of a strategy and direction for the coordination of its activities is seen as the primary direction for the community to establish to move towards building an effective and successful timing infrastructure.

This roadmap sets out some initial findings, ideas for development, and a proposal for a framework to establish an enduring talent and skills development ecosystem to support timing and associated technologies take these technologies from research findings to implementation and commercial exploitation.

**Current and future challenges**

The policy frameworks and infrastructure for skills and talent development already exists, to a greater or lesser extent, within the UK, housed within its world leading universities (Hubs and CDTs), industry, and government laboratories (DSTL, NPL).

The primary challenges that it faces are: Identifying clearly the needs and gaps (i.e. training requirements and capacity to deliver); ensuring that the talent development ecosystem is proactive and dynamic to ensure it meets rapidly evolving needs; and ensure an extensive and diverse pool of talent moves into the timing community. For these, and other yet to emerge, challenges to be met, it will be necessary to establish a coordinating engine to drive the system to deliver in a coherent and collaborative way.

Some significant challenges exist in attracting talent and ensuring there is an enduring and sustained supply of people into the system. The issues with this extend as far back as encouraging primary school age children to take up studying science and engineering, and for this to be maintained across a wide demographic as potential candidates move through the education system.

For any set of recommendations to be implemented successfully, it is suggested that these activities should focus on having an impact in the short-, medium- and long-term, thereby ensuring that we see successful outcomes in a timely way to inform and inspire future plans and actions. This coordinated activity also needs to be fully integrated across the main stakeholder groups, namely: Industry, Academia, Government and/or National Laboratories, Learned Societies, and Trade Associations.

The proposals set out above are derived to some extent from the findings set out in the UK Quantum Skills Taskforce report [4] alongside an exercise being carried out during 2025 that is surveying the timing community to answer 3 key questions, namely:

1. Should 'The National Timing Community engage in a collective effort to build an enduring Skills and Talent (S&T) development system"?
2. What are the drivers that enable participation in, and contributions to, the development of the timing skills and talent eco-system?
3. What are the organisational and personal blockers that could impact on the development and delivery of a skills and talent development system?

The overarching driver for this activity, namely the roadmap for the Talent and Skills development system is shown in Figure 1, with some of the key outcomes set out in the section below.

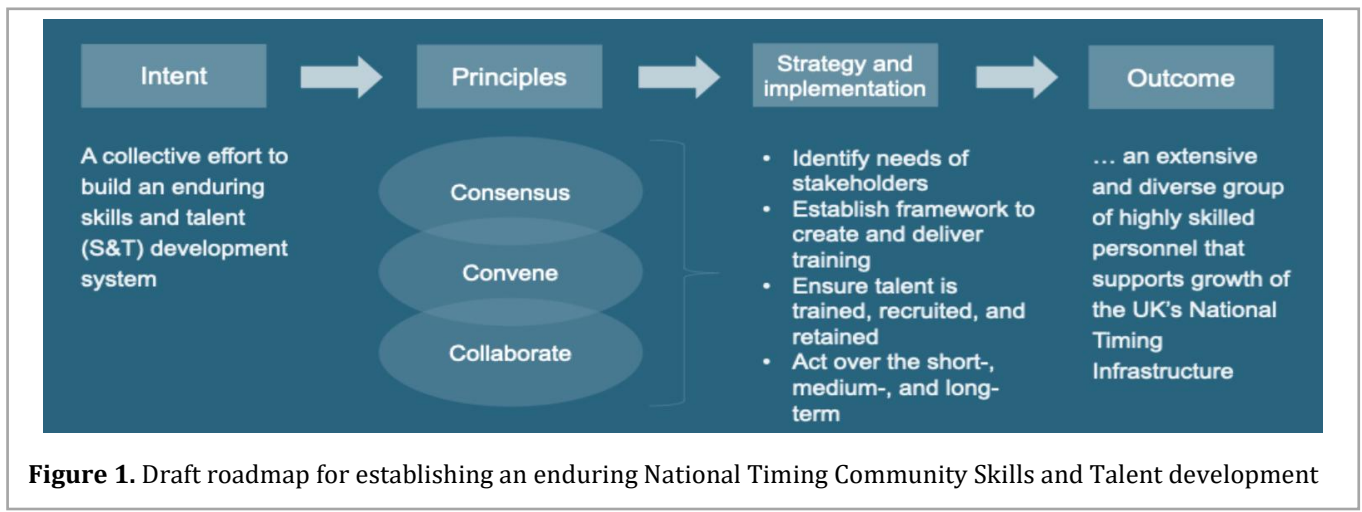

**Figure 1.** Draft roadmap for establishing an enduring National Timing Community Skills and Talent development

**Advances in skills coordination and development framework**

As stated above, extensive investment and effort is being made in the UKRI/EPSRC CDTs and Quantum Hubs to develop some critical elements of the talent and skills infrastructure for the PNT community. Allied to this, the work of the national laboratories such as Dstl and the National Physical Laboratory (through the National Timing Centre Programme) are examples of where a publicly funded national resources are providing strategic infrastructure for the advancement of timing technology. Furthermore, NPL's Postgraduate Institute [6] provides an environment where talent development across a broad range of skills is having an impact on developing the supply of capable postdoctoral researchers in a variety of technical areas that are essential to support the timing ecosystem.

The development of talented scientist and engineers also requires experience in real-world environments which can sometimes be difficult to access, but initiatives such as the STEM Futures scheme [7] – hosted by the Government Office for Science – provides a highly efficient mechanism to enable staff mobility for the purpose of training and development across a broad range of subject areas, amongst a wide range of academic, industrial and public bodies able to host secondees in highly specialised environments.

Within the Defence Science and Technology Laboratory (Dstl), the Cross-Cutting Timing Consortium (CCTC) has been established to primarily support provision of timing and time transfer technologies for defence and security sector. It is designed to contribute to the successful implementation of Distributed Resilience Timing/Time Transfer (DRT), the Resilient Enhanced Time Scale Infrastructure (RETSI) and applications that are dependent upon accurate, resilient timing in the following ways:

- Ensuring UK sovereign development of assured components for Ministry of Defence (MOD) DRT, RETSI and time distribution systems.
- Aiding awareness, coherence, and health of the UK's Timing & Time Transfer research landscape.
- Driving Suitably Qualified and Experienced Personnel development within Dstl, MOD, UK government, industry and academia.

This will be implemented, where appropriate and possible, by: Communicating requirements for timing, DRT & RETSI in Dstl and government; Identify calls for proposals; timing/time-transfer systems/component development; Highlighting relevant UK legislation; Communicate technical work in MOD, industry, academia and UK government; Identifying collaboration and secondment opportunities; and Promoting awareness of timing as an underpinning and critical capability

The exercise to survey the timing community and confirm the need for a coherent and collaborative approach to deliver a talent and skills development system, also asked for interested stakeholders to identify key drivers and blockers for this endeavour. Some initial outcomes are provided in Table 1 below.

**Table 1.** Extract of processed feedback from NPL/DSTL hosted timing community event held on 10 March 2025.

|  | **High popularity** | **Medium popularity** | **Niche but critical** |
|---|---|---|---|
| **What is needed?** | Career pathways and awareness | Funding and incentives | Policy, compliance and national security |
| **Why it matters?** | Lack of awareness cited as a barrier - sector needs a coherent, visible story | Many organisations can't invest due to cost/risk, especially SMEs | Timing is critical to national infrastructure, but poorly understood |
| **Suggested Action(s)** | Launch national awareness campaigns; central job hub; career videos in schools | Co-funded public/private training schemes; training tax incentives | Embed timing skills in national resilience frameworks; standards literacy across industry |

**Concluding remarks**

The proposal set out above provides some of the salient points that will support the development of a robust and enduring talent development structure for timing and PNT more generally and is set within the wider quantum skills landscape.

The proposed roadmap is predominantly driven by the context within which it has been framed, namely government strategy and policy directions and how this flows down to an infrastructure that has, and will continue to, receive significant investment. The criticality of coordination and coherence across the community to enable effective collaboration cannot be overstated. The current plan is time sensitive by definition as it is evolving and is required to be agile in the face of rapidly changing or evolving technologies, needs, and capabilities. They key conclusion, therefore, to ensure we have a strong community and enduring talent and skills development ecosystem is that in terms of action, "the time is now!"

**Acknowledgements**

The authors would like to acknowledge the support of NPL (National Timing Centre Programme), Dstl (StratCom), The Royal Institute of Navigation, The Quantum Enabled PNT Hub, for the support and enabling the surveys at several community events. Michael Lingard (NPL) for the development of the roadmap framework.

# Microcombs for Portable Optical Clocks

**Jonathan Silver[1]**

[1] National Physical Laboratory, Teddington, United Kingdom

E-mail: jonathan.silver@npl.co.uk

**Status**

Optical atomic clocks [1] are the state of the art in timekeeping, with systematic uncertainties and stabilities at the level of 1 part in $10^{18}$ or below demonstrated with averaging times of just a few hours. By comparison, caesium fountain clocks, the leading microwave atomic clocks that are used in national metrology institutes around the world to generate Coordinated Universal Time (UTC) based on the current definition of the second, typically have uncertainties and stabilities of a few parts in $10^{16}$ with averaging times of weeks [2]. Since optical clocks are able to measure time so much more accurately, work is underway to redefine the second in the coming years based on one or more optical atomic transitions [3]. This level of accuracy enables entirely new applications from Earth monitoring via measurement of the gravitational potential with centimetre height resolution to detection of drifts or local variation of fundamental constants caused for example by dark matter. Furthermore, portable optical clocks of various sizes are being developed that deliver better stability performance than microwave clocks of comparable size, weight and power consumption (SWaP) [4], which could eventually replace these in their current applications and unlock many new ones in areas such as telecommunications and positioning, navigation and timing (PNT).

A central component in any optical clock is the frequency comb [5]. These enable coherent conversion of frequencies both between the optical and microwave/RF domains and across the optical domain, allowing the counting of optical clock frequencies via down-conversion to RF as well as direct optical frequency ratio measurements over fibre, both locally and remotely via telecom fibre links. For a comb to be used for frequency ratio measurements, it must be self-referenced, meaning that the carrier-envelope offset frequency is detected and/or locked. This is usually achieved by frequency-doubling the low-frequency end of an octave-spanning comb and taking the beat note of this with the other end of the comb. Current commercial self-referenced combs are based on femtosecond modelocked lasers that are broadened to an octave in highly nonlinear fibres or waveguides. Although these systems perform well and can be reasonably robust, they range from a few litres to about a cubic metre in total volume, are hand-built and typically cost at least £100,000, which limits their uses. Having a mass-producible self-referenced comb with a much smaller form factor is therefore key to realising compact optical clocks as well as other applications of coherent frequency conversion such as optically-derived ultrastable microwave sources for radar.

In recent years, microresonator-based frequency combs, or microcombs for short [6], have emerged as a very promising route to ultracompact combs. Based on ultrahigh-Q whispering-gallery-mode microresonators that convert a monochromatic pump laser into an entire comb via their intrinsic Kerr optical nonlinearity, they have been demostrated in many types of resonator including chip-based waveguide microrings created from a range of material platforms [7]. Various different types of phase-locked microcombs have been produced including single-soliton, multi-soliton and soliton crystal combs in resonators with anomalous dispersion as well as dark solitons, platicons and switching waves in the normal-dispersion regime [8]. In contrast to modelocked-laser-based combs, which typically have repetition rates of hundreds of MHz, microcombs usually have spacings in the range 10 GHz to 1 THz due to their much smaller cavity sizes. With the rapid recent pace of development of microphotonics fabrication processes, there is now a clear pathway to realising mass-producible chip-based self-referenced frequency combs, as outlined in the following sections.

**Current and future challenges**

In the last decade or so, silicon nitride has emerged as the leading platform for integrated microcombs in the near-infrared, due to its favourable dispersion properties, CMOS-compatible fabrication processes, relatively high Kerr nonlinearity (10 times that of silica) and the ultrahigh Q factors of over 10 million that have been achieved for high-confinement ring resonators [9]. The high repetition rates of microcombs mean that external broadening to an octave in a single pass through a nonlinear fibre or waveguide requires amplification to average powers of several watts [10]. As this is not currently feasible on chip, the octave-spanning comb must be generated inside a microresonator if the entire optical circuit is to be on chip. Octave-spanning soliton microcombs with a 1 THz repetition rate have been produced in silicon nitride microring resonators with as low as 40 mW of pump power on chip [11] and with higher pump powers at repetition rates down to around 200 GHz. In these studies, the resonator's dispersion was engineered via the waveguide width and thickness to allow broadband comb generation and place dispersive wave peaks in the spectrum an octave apart from each other to increase the power available for self-referencing.

Octave-spanning microcombs with repetition rates below 200 GHz are difficult to produce in phase-locked states, as well as requiring pump powers on the level of 1 W or above due to the larger resonator mode volume, and so are impractical. On the other hand, the limited availability of microwave components above 200 GHz poses a challenge to measuring such high repetition rates, which can be overcome in a number of ways. One is to use a microcomb with a gigahertz repetition rate to subdivide the repetition rate of the octave-spanning comb [12], but more recently a Vernier technique has been demonstrated using two terahertz-spacing combs with slightly different spacings, which has the added advantage that it allows terahertz carrier-envelope offset frequencies to be measured at the same time [13].

Another challenge is the integration of the pump laser with the microcomb due to the fact that diode lasers are based on III/V semiconductors which cannot be directly grown on a silicon-based platform due to the lattice constant mismatch. However, much progress has been made recently with heterogeneous integration techniques such as butt-coupling, flip-chip bonding and wafer bonding, and these have been used to integrate distributed feedback (DFB) lasers and reflective semiconductor optical amplifiers (RSOAs) with silicon nitride photonic integrated circuits (PICs) for on-chip pumping of microcombs, such as in [14]. In this and other demonstrations, the pump laser was self-injection-locked to a resonant Rayleigh back-reflection from the ultrahigh-Q resonator that generates the comb. This meant that the laser-resonator detuning was hardly affected by thermal nonlinearities, circumventing the difficulties encountered with entering the single-soliton comb state when using an external pump laser [8] and instead allowing simple turnkey access to it upon switching on the pump current.

Single-bright-soliton combs have certain advantages compared to other types of phase-locked microcombs, including stability, reproducibility, the potential for very broad spectra and the ability to generate strong beat note signals such as the repetition rate due to the alignment of the comb lines' phases. However, they also have one notable disadvantage when generated in a single resonator pumped with continuous-wave (CW) light, namely their low pump-to-comb power conversion efficiency, which scales as the inverse of the number of comb lines within the 3 dB bandwidth of the spectral envelope [15]. Several solutions to this have been demonstrated including pulse pumping, multi-soliton and soliton crystal combs, dark solitons, laser cavity solitons and schemes involving dual resonators and photonic crystal resonators [16]. In one case, an auxiliary resonator was used to selectively shift the pump resonance frequency, allowing the production of a single-bright-soliton comb with more than 50% conversion efficiency from a CW pump, a result that was later improved upon by achieving the same effect using a photonic crystal resonator.

One potential disadvantage of microcombs compared with modelocked-laser-based combs for optical frequency metrology applications is their high thermorefractive noise due to their small mode volume, which adds phase noise to their repetition rate and carrier-envelope-offset frequencies. Although this noise can be effectively suppressed by applying feedback to the pump laser and resonator microheater currents, it is still difficult to reach the stabilities possible with modelocked-laser-based combs using this method. However, a remarkably good solution to this is to injection-lock a comb line far from the pump to a second laser coupled into the resonator [17].

**Advances in science and technology to meet challenges**

Although the challenges of ultrahigh-Q microring resonator fabrication, on-chip comb generation, pump laser integration, octave-spanning comb generation, terahertz repetition rate measurement, efficient comb generation and noise suppression have all separately been solved, there is still work to be done to combine these all to produce a low-noise, power-efficient, octave-spanning microcomb on a chip. Furthermore, second-harmonic generation (SHG) must also be performed on the same chip to self-reference the comb and potentially also to frequency-double specific comb lines to produce visible optical clock wavelengths, since silicon nitride microcombs generally occupy the 1-2 μm wavelength range. A promising route to achieving this is to use periodically-poled thin-film lithium niobate (PPTFLN) ring resonators, which can achieve extremely high SHG efficiencies through a combination of high confinement, resonant enhancement and artificial phase-matching [18]. Integration of lithium niobate onto silicon nitride is currently an area of active research, with recent progress made on both transfer printing of devices and wafer bonding of membranes. There has also been a recent demonstration of two-colour soliton microcomb generation in PPTFLN ring resonators [19], which suggests the possibility of generating a self-referenced comb directly in a PPTFLN resonator by making the fundamental and second-harmonic combs broad enough that their spectra overlap through careful dispersion engineering and increasing the pump power. Finally, high-speed photodiodes must also be integrated onto the chip for beat note detection, although this does not present a significant additional challenge as both the waveguide-based photodetectors and the techniques for their heterogeneous integration have already been developed.

**Concluding remarks**

Microcombs offer an extremely promising route to achieving a mass-producible self-referenced frequency comb on a chip, which could be truly revolutionary, enabling portable, compact or even fully chip-based optical clocks [20] as well as ultrastable microwave sources and other technologies. Most of the challenges to realising chip-based self-referenced combs have already been individually addressed, and with the current rate of progress in microphotonics fabrication techniques, it is highly possible that they will be produced in the next few years.

**Acknowledgements**

We acknowledge the support of the UK Government Department for Science, Innovation and Technology through the UK National Quantum Technologies Programme.